# When do complex transport dynamics arise in natural groundwater systems?

**J. Wu[1], D. R. Lester[1], M. G. Trefry[2], G. Metcalfe[3]**

[1] School of Engineering, RMIT, Melbourne, Victoria, Australia.

[2] Independent Researcher, Floreat, Western Australia, Australia.

[3] School of Engineering, Swinburne University of Technology, Hawthorn, Victoria, Australia

Corresponding author: D.R. Lester (daniel.lester@rmit.edu.au)

**Key Points:**

- Transient Darcy flows can generate complex transport dynamics in heterogeneous compressible aquifers.

- This complex transport can trap dispersing solutes for many years.

- Global tidal maps indicate widespread potential for complex transport dynamics in coastal zones.

## Abstract

In a recent paper (Trefry et al., 2019) we showed that the interplay of aquifer heterogeneity and poroelasticity can produce complex transport in tidally forced aquifers, with significant implications for solute transport, mixing and reaction. However, what was unknown was how broadly these transport dynamics can arise in natural groundwater systems, and how these dynamics depend upon the aquifer properties, tidal and regional flow characteristics. In this study we answer these questions through parametric studies of these governing properties. We uncover the mechanisms that govern complex transport dynamics and the bifurcations between transport structures that depend upon changes in the governing parameters, and we determine the propensity for complex dynamics to occur in natural aquifer systems. These results clearly demonstrate that complex transport structures and dynamics may arise in natural tidally forced aquifers around the world, producing solute transport and mixing behaviour that is very different to that of the conventional Darcy flow picture.

## 1. Introduction

Many natural groundwater systems involve discharge boundaries at rivers, lakes and coasts where subsurface aquifer fluids interact with fresh or saline fluids at the discharge boundary, leading to strong chemical and nutrient gradients and highly localized hotspots of chemical and biological activity. Invariably, these discharge boundaries have transient forcing via e.g. tidal action, seasonal variations, barometric effects or long-range climatic cycles, which interact with the regional aquifer flow to generate complicated mixing phenomena near the discharge boundary. One important class of such systems is coastal aquifers, where tidal forcing is critical to both saltwater intrusion into the aquifer and a wide range of physical and chemical processes. Whilst much attention has been focused on the role of density-driven mixing in the Ghyben-Herzberg zone, the interplay of tidal forcing with regional groundwater flow can also generate complex mixing phenomena that includes chaotic mixing and trapping of solutes. Whilst several studies (Li et al., 2004; Pool et al., 2015; Robinson et al., 2009) have considered the interplay of tidal forcing and regional groundwater flow on solute transport, to date no field studies have been attempted to explicitly resolve the kind of complex transport behaviour that is the subject of this paper. Similarly, transient forcing of lacustrine and riverine aquifers can generate complex mixing phenomena between water bodies of different bio-geo-chemical origin, leading to highly localized concentration gradients and complex transport phenomena. The transient nature of these complex flow dynamics means they can be difficult to visualize and understand (even in two-dimensional idealizations), and specialized mathematical tools and techniques are required to elucidate the impacts on solute transport and associated phenomena such as chemical reaction or biological activity.

Recently, in Trefry et al. (2019) (called TLMW henceforward for brevity) we uncovered mechanisms that can lead to complex mixing in periodically forced Darcy groundwater systems, where the interplay of tidal forcing, aquifer compressibility and heterogeneity caused important changes in the flow structure and transport dynamics. Using a highly simplified, vertically averaged two-dimensional (2D) aquifer model, we showed that the interplay of these properties can result in localized *flow reversal* (where the groundwater flow vector makes a sweep through all orientations) over the periodic forcing cycle. Such changes in flow orientation reorganize the effective Darcy flow topology away from the conventional flow structures of steady Darcy flow. We term these changes *topological bifurcations* of the aquifer transport structure and, as shown below (Figure 2), these bifurcations fundamentally change the organization of fluid trajectories in the aquifer and lead to transport dynamics that are radically different to those of steady Darcy flow. Unlike steady 2D flows, where the transport dynamics can be directly visualized from the streamlines of the flow, transient 2D flows require different methods to understand the transport structure. The simplest way to visualize this transport structure is via a *Poincaré section* (Aref, 1984; Ottino, 1989), which records fluid particle positions at integer multiples of the forcing periods, effectively filtering out the "fast" intra-period particles trajectories, leaving only the "slow" time-averaged transport. TLMW showed that flow reversal can lead to two major changes in the slow, time-averaged transport structure. First, flow reversal can lead to a change in the effective transport structure of natural aquifers from open 1D particle trajectories without fixed points (i.e. similar to streamlines for steady Darcy flow), to closed 1D particle trajectories with fixed points. Second, even stronger flow reversal can then lead to a bifurcation of the closed flow structure to *chaotic* particle trajectories, where the closed 1D particle trajectories (in the periodic sense) break up into a 2D scattering of particle locations that lead to rapid mixing and strongly anomalous transport. We collectively term these profound changes in the transport structure of these flows to closed and chaotic particle orbits as *complex transport dynamics* in the aquifer. Complex transport dynamics determine the mixing, dispersion and residence time characteristics of these flows, with significant implications for solute transport, chemical reactions and biological activity (Heiss et al., 2017; Liu et al., 2018; Slomp & Van Cappellen, 2004; Tél et al., 2005).

Such fundamental changes in the transport structure of tidal aquifers are governed by a small number of dimensionless parameters which are related to the aquifer properties, tidal forcing, and regional flow characteristics. In TLMW, we showed that complex transport dynamics can arise in coastal aquifers with physical properties similar to those of several naturally occurring systems.

However, as in that study we considered a single set of governing parameters, the transport and mixing behaviour of these systems across the full dimensionless parameter space is presently unknown. This question is relevant to understanding the nature and ubiquity of complex transport dynamics in tidally forced aquifers and implications for physical processes such as mixing, solute dispersion, seawater intrusion, pollutant transport, chemical reactions and biological activity. Therefore, here we scan the parameter space and examine the propensity for complex transport dynamics to occur in field settings.

To address this question, we perform a range of simulations using the periodically forced 2D aquifer model introduced in TLMW to resolve the aquifer transport dynamics over the relevant parameter space. Precisely defined below, these parameters include the strength of tidal forcing relative to that of the regional flow, relative compressibility of the aquifer, and the attenuation of the tidal fluctuation into the aquifer, as well as the statistical parameters of the random hydraulic conductivity field. From these results we determine the key transport characteristics over the flow parameter space and develop a phase diagram for the different transport structure topologies (i.e. open, closed or chaotic). We then use these results to develop mechanistic arguments for how and why transport varies over the parameter space, and how these governing parameters for natural systems relate to these regions of augmented transport. Whilst solute transport is governed by the interplay of fluid advection and molecular diffusion or hydrodynamic dispersion, in this study we focus on the purely advective dynamics imparted by transient forcing of the aquifer. This approach, which uncovers the *Lagrangian kinematics* of these flows, facilitates a clearer visualization and understanding of the transport structures that govern solute transport in periodically forced aquifers. Finally, we estimate the impacts of diffusion upon solute transport and the propensity for complex transport to occur in natural coastal aquifers.

The purpose of this study is to show when, how and why this previously unresolved mechanism for complex transport can arise , and to point out potential implications for solute transport, dispersion and mixing. Because this complex transport mechanism is the primary focus of this study, we do not aim to present complete predictions of solute transport from a fully resolved 3D aquifer model. Rather, we limit consideration to an idealised 2D aquifer model with a multi-Gaussian conductivity field that does not include e.g. density-driven effects. Whilst we are aware that 3D coastal aquifers and different conductivity models can generate much more complicated flow and transport phenomena (Pool et al., 2011), the reasons for using such a 2D model are: First, we shall show that the possible mechanisms and kinematics of these transport dynamics in 2D are already quite complex, and so we need to fully understand these dynamics before consideration of 3D flow and transport. Second, density-driven flows in coastal aquifers often only penetrate relatively short distances from the oceanic interface, whereas, strong tidal signals may penetrate several kilometres inland. In Section 2.2 we shall show that for many common aquifers the saline wedge only penetrates a fraction of the distance of the tidal signal. As such, many of the transport structures resolved in this study are not influenced by density driven flows at coastal boundaries. Third, as more complex transport is expected in actual 3D aquifers, we employ the idealized 2D model under the assumption that predictions of this model provide a conservative estimate for the generation of complex transport structures in 3D aquifers. This corresponds with general observations in spatially extended dynamical systems that 3D systems tend to transition to chaotic dynamics more readily than their 2D analogues, as the additional spatial dimension admits a richer set of dynamics. Certainly, there exist many interesting and important research questions regarding (i) the manifestation of these transport dynamics in 3D aquifers, (ii) interactions between this transport mechanism and density-driven mixing, (iii) the impact of different conductivity structures, (iv) the impact of solute diffusion and dispersion, (v) the impact of biogeochemical reactivity, and combinations thereof. In subsequent studies we will consider the interplay of advective transport, solute diffusion and dispersion, and multi-modal forcing.

The remainder of this paper is organized as follows. In Section 2 we review the model for flow in a heterogeneous tidally forced aquifer and define the key dimensionless parameters. We then examine the governing tidal mixing mechanisms in Section 3 and topological bifurcations of the transport structure (Lagrangian kinematics) over the aquifer control parameter space in Section 4. In Section 5 we estimate the impact of complex transport upon solute diffusion and dispersion, and in Section 6 we consider the propensity for complex transport to occur in natural systems before conclusions are made in Section 7.

## 2. Model description

### 2.1. Governing equations

To begin we briefly review the model for flow in a heterogeneous tidally forced aquifer introduced in TLMW as this will serve as a basis for our calculations. Following conventional approaches (Bear, 1972; Coussy, 2004), we consider the fluid and solid phases within the aquifer to be incompressible, yet the solid matrix may undergo differential compression via the migration of solid particles due to gradients in the pore pressure field. We consider a vertically averaged 2D linear groundwater model (Figure 1) that is driven by a constant regional flow gradient $J$ from right to left in an aquifer with a spatially heterogeneous conductivity field $K$. We note that the *inland* (right, $x = L$) and *no flow* (bottom, $y = 0$; top, $y = L$) boundaries do not correspond to physical features of the aquifer, rather these boundaries are the physical extent of the computational model. Conversely, the *tidal boundary* (left, $x = 0$) is a discharge boundary to an open flow system such as oceanic, riverine or lacustrine flow, where tidal, seasonal, barometric or climatic variations give rise to a temporally fluctuating flow potential. The interaction between this oscillating potential at the tidal boundary and the regional flow gradient can give rise to complex transport dynamics within the aquifer.

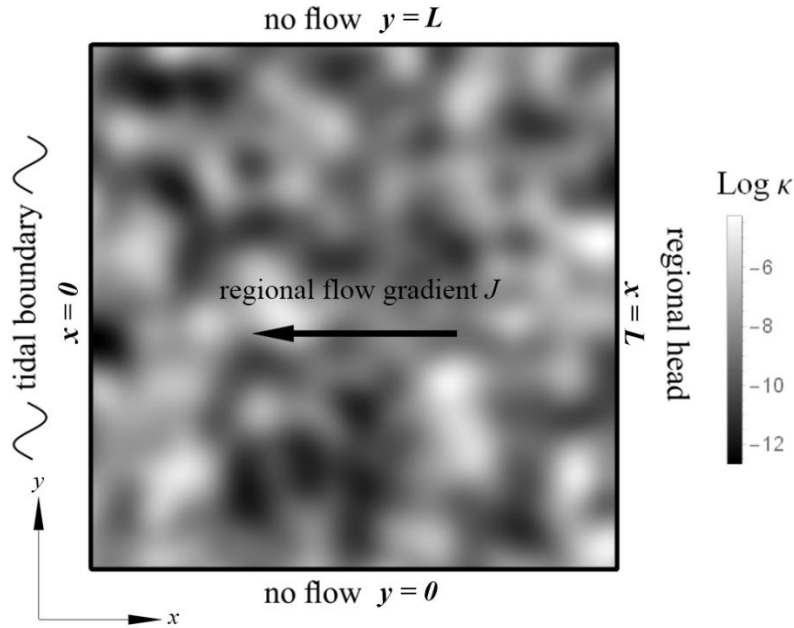

**Figure 1.** Schematic of the model 2D aquifer domain with boundary conditions and spatial heterogeneity of $K$.

In this linear groundwater model, the Darcy flux **q** and aquifer porosity $\varphi$ are governed respectively by the Darcy flow and linearized porosity-storativity equations (Bear, 1972; Coussy, 2004)

$$\mathbf{q} = -K\nabla h, \tag{1}$$

$$\varphi(h) = \varphi_{\mathrm{ref}} + S(h - h_{\mathrm{ref}}), \tag{2}$$

where $h$ is the pressure head (with reference value $h_{ref}$), $\varphi$ the aquifer porosity (with reference value $\varphi_{ref}$), $S$ is the constant aquifer storativity that arises from the migration of solid particles in response to gradients in $h$. It is important to note that whilst the fluid and solids phases are regarded as individually incompressible, such migration leads to differential compression (or consolidation) of the solids phase which is henceforth referred to as aquifer compressibility. In equation (1), $K(\mathbf{x})$ is the heterogeneous isotropic saturated hydraulic conductivity which varies with the physical space coordinate $\mathbf{x} = (x, y)$. Equations (1) and (2) may be combined via the continuity equation

$$\frac{\partial \varphi}{\partial t} + \nabla \cdot \mathbf{q} = 0 \qquad (3)$$

into the linear groundwater flow equation in terms of the pressure head $h$

$$S \frac{\partial h}{\partial t} = \nabla \cdot (K \nabla h), \qquad (4)$$

subject to the no flow boundary conditions at $y = 0$ and $y = L$, fixed head condition at the inland boundary $x = L$ and periodically fluctuating head at the tidal boundary $x = 0$,

$$\frac{\partial h}{\partial y}|_{y=0} = \frac{\partial h}{\partial y}|_{y=L} = 0, \qquad h(L, y, t) = JL, \qquad h(0, y, t) = g(t), \qquad (5)$$

where $g(t) = g_p e^{i\omega t}$ is the periodic tidal forcing function. Whilst most tidal forcings are multi-modal, for simplicity we consider tidal forcing $g(t)$ via a single Fourier mode with constant forcing frequency $\omega$ (and period $P = 2\pi/\omega$) and amplitude $g_p$. We anticipate that multi-modal forcing shall act to further increase the complexity of transport dynamics in transiently forced aquifers, and this expectation is supported by preliminary results from current studies on these systems. Under unimodal forcing, equation (4) can be separated into steady $h_s$ and periodic $h_p$ components of the pressure head $h(\mathbf{x}, t) = h_s(\mathbf{x}) + h_p(\mathbf{x}) e^{i\omega t}$ which individually satisfy

$$\nabla \cdot (K \nabla h_s(\mathbf{x})) = 0, \qquad \nabla \cdot \left( K \nabla h_p(\mathbf{x}) \right) - i\omega S h_p(\mathbf{x}) = 0, \qquad (6)$$

and the steady and periodic boundary conditions

$$h_s(L, y) = JL, \qquad h_s(0, y) = 0, \qquad h_p(L, y) = 0, \qquad h_p(0, y) = g_p. \tag{7}$$

From the continuity equation (3) the porosity $\varphi$ may also be decomposed into steady $\varphi_s$ and periodic $\varphi_p$ components as

$$\varphi_s(\mathbf{x}) = \varphi_{\text{ref}} + S(h_s(\mathbf{x}) - h_{\text{ref}}), \qquad \varphi_p(\mathbf{x}) e^{i\omega t} = Sh_p(\mathbf{x}) e^{i\omega t}. \tag{8}$$

To visualize the transport structures in the periodically forced aquifer, we heavily utilize particle tracking, where the position $\mathbf{x}$ of the fluid particle evolves with time $t$ as per the advection equation

$$\frac{d\mathbf{x}}{dt} = \mathbf{v}(\mathbf{x}, t) = \frac{\mathbf{q}(\mathbf{x}, t)}{\varphi(\mathbf{x}, t)} \tag{9}$$

where $\mathbf{v}(\mathbf{x}, t)$ is fluid velocity defined as the Darcy flux normalised by the local porosity. TLMW detailed the computational method to accurately compute $\mathbf{v}(\mathbf{x}, t)$ to ensure exact satisfaction of the continuity equation (3). In the Supplementary Material (S1) we describe a method to rapidly solve (9) for large numbers of fluid particles.

The Darcy flux $\mathbf{q}(\mathbf{x}, t) = (q_x, q_y)$ at a fixed point $\mathbf{x} = (x, y)$ within the aquifer has steady and periodic components

$$\mathbf{q}(\mathbf{x}, t) = \mathbf{q}_s(\mathbf{x}) + \mathbf{q}_p(\mathbf{x}, t) \equiv \mathbf{q}_s(\mathbf{x}) + \mathbf{q}_p(\mathbf{x}) e^{i\omega t}, \tag{10}$$

which lead to elliptical Darcy flux orbits over a forcing period $P = 2\pi/\omega$ if $||\mathbf{q}_p(\mathbf{x})|| > 0$. *Flow reversal* occurs at a given point $\mathbf{x}$ in the aquifer during the forcing period if the magnitude of the periodic flux at any point during the period is greater than that of the steady flux, i.e. $||\mathbf{q}_p(\mathbf{x}, t)|| > ||\mathbf{q}_s(\mathbf{x}, t)||$, resulting in a *canonical* flux ellipse (see TLMW for details).

This conventional linear groundwater flow model is the basis for study of the interplay of tidal forcing, non-zero matrix compressibility and aquifer heterogeneity, leading to complex transport dynamics in periodically forced aquifers. As the advection equation (9) represents a transform from the Eulerian to the Lagrangian frame, we shall make extensive use of particle tracking via (9) to study the Lagrangian kinematics and transport dynamics of periodically forced aquifers.

## 2.2. Key dimensionless parameters

TLMW identified a set of dimensionless parameters that characterize the solutions to the 2D tidal forcing problem introduced above. In addition to these dynamical parameters, there also exist a number of statistical parameters which characterize the heterogeneous hydraulic conductivity field $K(\mathbf{x})$. Whilst generation of complex transport dynamics in periodically forced aquifers requires a heterogeneous hydraulic conductivity field to generate flow reversal, we conjecture that this phenomenon persists regardless of the particular statistical autocorrelation model used for the conductivity field. In this study we consider a random log-Gaussian conductivity field $K$ that is statistically determined by its mean conductivity $K_{\text{eff}}$, log-variance $\sigma_{\log K}^2$ and correlation length $\lambda$, and so the independent set of physical parameters that characterize the statistical conductivity field is then $\chi \equiv (K_{\text{eff}}, \sigma_{\log K}^2, \lambda)$.

**Table 1.** Dimensionless parameters for the tidal forcing problem; see TLMW for details.

| Dimensionless Parameter | Name | Definition | Range | Physical Motivation |
|---|---|---|---|---|
| *Dynamical Parameter Set $\mathcal{Q}$* | | | | |
| $\mathcal{T}$ | Townley number | $\mathcal{T} \equiv \dfrac{L^2 S \omega}{K_{\text{eff}}}$ | $[0, \infty)$ | Ratio of diffusive and tidal forcing time scales |
| $\mathcal{G}$ | Tidal strength | $\mathcal{G} \equiv \dfrac{g_p}{JL}$ | $[0, \infty)$ | Ratio of tidal amplitude to inland head |
| $\mathcal{C}$ | Tidal compression ratio | $\mathcal{C} \equiv \dfrac{\Delta \varphi}{\varphi_{\text{ref}}} = \dfrac{S g_p}{\varphi_{\text{ref}}}$ | $[0, 1]$ | Maximum relative change in porosity due to tidal variation |

| *Heterogeneity Parameter Set* $\chi$ | | | | |
|---|---|---|---|---|
| $\mathcal{H}_t$ | Temporal character | $\mathcal{H}_t \equiv \dfrac{\lambda \varphi_{\text{ref}} \, \omega}{2\pi \, K_{\text{eff}} |J|}$ | $[0, \infty)$ | Ratio of Darcy drift and tidal time scales |
| $\mathcal{H}_x$ | Spatial character | $\mathcal{H}_x \equiv x_{\text{taz}} / \lambda$ | $[0, L/\lambda)$ | Number of correlation scales in the tidally active zone $x_{\text{taz}}$ |
| $\mathcal{G} \, \sigma_{\log K}^2$ | Vorticity number / Flow reversal number | $\mathcal{G} \, \sigma_{\log K}^2$ | $[0, \infty)$ | Density of flow reversals in the tidally active zone |

In this paper we will study advective flow in the linear groundwater model over the tidal flow parameter space $\mathcal{Q} \equiv \mathcal{T} \times \mathcal{G} \times \mathcal{C}$, identify regions of anomalous transport and chaotic advection, and comment on how $\mathcal{Q}$ and $\chi$ together control these topological transitions and the underlying physical mechanisms. Chaotic advection arises when fluid particles undergo chaotic orbits, leading to exponential stretching of fluid elements with time, rapid mixing and augmented transport (Ottino, 1989; Trefry et al., 2019). Table 1 summarises the various dimensionless dynamical and heterogeneity parameters and provides a brief physical motivation for each. The tidally active zone ($0 < x < x_{\text{taz}}$) corresponds to the region of the aquifer that is subject to significant forcing signals from the tidal boundary ($x=0$). In TLMW we show that the extent of the tidally active zone ($x_{\text{taz}}$) in the aquifer is a function of the parameters $\mathcal{T}$ and $\mathcal{G}$. In S7 of the supplementary material we show that for most coastal aquifers, the penetration distance of the saline wedge is only a small fraction of $x_{\text{taz}}$, hence there typically exists a significant portion of coastal aquifers that are not subject to saline density currents but are still influenced by the tidal boundary.

## 3. Transport structures of periodically forced aquifers

The complex transport structures which arise in periodically forced aquifers require specialized tools for their visualization and understanding, which may be unfamiliar to researchers in groundwater hydrology. As shall be shown, we observe three main types of transport structures (open, closed, chaotic) in periodically forced aquifers over the flow parameter space $\mathcal{Q} = \mathcal{T} \times \mathcal{G} \times \mathcal{C}$. In this Section we briefly review these visualization tools, classify and describe these transport structures and show how they change with $\mathcal{T}, \mathcal{G}, \mathcal{C}$.

### 3.1. Visualisation of transport structures and Lagrangian kinematics

In the absence of sources and sinks, steady Darcy flow is typified by open streamlines and an absence of stagnation points (Bear, 1972), leading to slow mixing and limited transport dynamics (Dentz et al., 2016). Conversely, unsteady Darcy flows can break these topological constraints due to transient switching of streamlines (Lester et al., 2009; 2010; Metcalfe et al., 2010a; Trefry et al., 2012), leading to a much richer set of possible transport structures. Particle trajectories in transient flows can become very complicated, so direct plotting does not provide much insight. To circumvent this problem, we use the *Poincaré section* (Aref, 1984; Ottino, 1989) to resolve the transport structure (or Lagrangian kinematics) of the flow by advecting (via equation (9)) a large number of fluid particles and recording the particle locations at each forcing period $P$. This filters out the "fast" oscillatory intra-period particle trajectories, leaving only the "slow" net particle motion. The Poincaré section can contain *regular*, non-chaotic regions which are comprised of coherent 1D particle trajectories that are non-mixing (in the ergodic sense), and distinct *chaotic* regions which are comprised of a seemingly random particle distributions which are formed as particles undertake chaotic, space-filling orbits. Chaotic regions are associated with localized strong (exponential) fluid stretching and rapid mixing, whereas the coherent trajectories in regular regions form barriers to transport.

We shall also make extensive use of Residence Time Distributions (RTDs) to characterize transport in the periodically forced aquifers. Although particle trapping is not possible in steady

Darcy flow, closed fluid orbits are possible in transient Darcy flow, leading to infinite residence times. These trapped regions are not necessarily chaotic, they can also be regular with closed streamlines. In this study we use two different particle injection protocols to generate RTDs and Poincaré sections: (i) inland boundary injection, where tracer particles are released in a flux-weighted manner along the inland boundary ($x = L$), and (ii) distributed injection, where particles are released uniformly over the entire aquifer domain. Distributed injection resolves transport structures that are not accessible to particles released along the inland boundary, whereas inland boundary injection highlights regions of the flow that are not accessible to these particles.

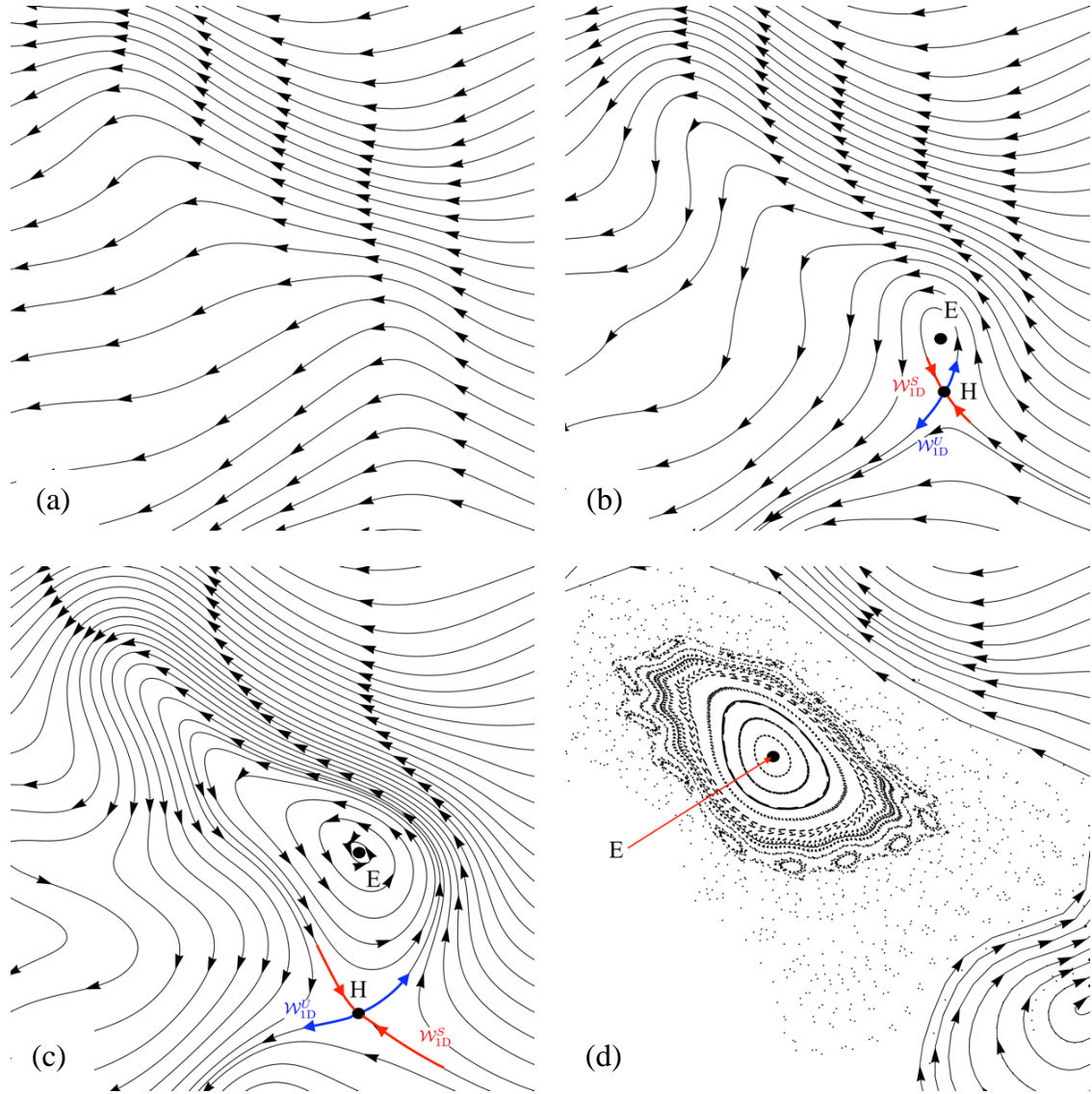

**Figure 2.** Bifurcation of transport structure types with increasing aquifer compressibility $\mathcal{C}$ with $\mathcal{T} = 10$, $\mathcal{G} = 100$ in the region (x, y) = (0.3, 0.5) × (0.75, 0.95) of the aquifer domain. Examples of (a) open ($\mathcal{C} = 0.002 < \mathcal{C}_1$), (b) closed ($\mathcal{C}_1 < \mathcal{C} = 0.005 < \mathcal{C}_2$), (c) closed ($\mathcal{C}_1 < \mathcal{C} = 0.1 < \mathcal{C}_2$) and (d) chaotic ($\mathcal{C}_2 < \mathcal{C} = 0.2$) transport structures in periodically forced aquifers. Arrows denote propagation of tracer particles along 1D trajectories, scattered points denote incoherent, chaotic motion of tracer particles. Dots denote elliptic (E) or hyperbolic (H) points, blue and red arrows denote unstable and stable manifolds. See text for more details.

## 3.2. Characteristic transport structures in periodically forced aquifers

Figure 2 shows the Poincaré sections for the three main types of aquifer transport structures (open, closed, chaotic) found in this study. Below we describe the features and dynamics of these structures, implications for transport and how they change with the flow control parameters. In this figure we have selected cases by only varying $\mathcal{C}$, but similar transitions are observed by varying the other physical parameters.

### 3.2.1. Open transport structures

As shown in Figure 2a, for small values of the relative compressibility $\mathcal{C}$, the aquifer Poincaré section is comprised only of regular and open (time-averaged) particle trajectories that resemble open streamlines of steady Darcy flow. Although particles undergo cyclical motion as they move through the aquifer, their basic transport structure consists of ordered and coherent 1D particle paths with no trapped regions or fixed points. Whilst these trajectories are displayed as continuous lines in Figure 2a, typically they would manifest as a series of discrete points in the Poincaré section. Such representation is not possible for the chaotic regions (as regular trajectories do not exist), hence chaotic regions are displayed using discrete points in Figure 2.

### 3.2.2. Closed transport structures

Above a critical value (termed $\mathcal{C}_1$) of the relative compressibility $\mathcal{C}$ (Figure 2b), the aquifer Poincaré section appears to undergo a *topological bifurcation* from open trajectories to particle trajectories that are now closed. This represents a fundamental change in the transport structure as fluid elements can neither enter nor leave this closed region, and so trapped fluid elements have essentially infinite residence times. In addition to these closed regions, *periodic points* of the flow also arise (where fluid particles return to the same position after one ($k = 1$) or more ($k > 1$) flow periods) which then manifest as fixed ($k = 1$) or periodic ($k > 1$) points in the Poincaré section. As shown in Figure 2, these periodic points may be classified as *elliptic* (E) or *hyperbolic* (H) points, based on their local transport structure. Elliptic points (E) involve a net rotation of the local fluid and are located in the centre of closed flow regions. Hyperbolic points (H) involve a net local saddle flow and are often found at the separatrices between closed and open transport structures. Fluid elements near hyperbolic points undergo exponential fluid deformation over each flow period, with stretching (blue) and contracting (red) particle trajectories shown in Figure 2c that are respectively termed the *unstable* ($W_{1D}^U$) and *stable manifolds* ($W_{1D}^S$) of the flow. These manifolds play a critical role in organizing transport. If these manifolds connect smoothly (Figure 2b, c), then the associated exponential stretching cancels out and regular particle trajectories result, but more complex transport dynamics arise if they intersect transversely.

### 3.2.3. Chaotic transport structures

As shown in Figure 2d, further increases in $\mathcal{C}$ beyond a second critical value $\mathcal{C}_2$ (where $\mathcal{C}_2 > \mathcal{C}_1$) leads to breakup of regular, coherent 1D particle trajectories into chaotic, space-filling particle trajectories that manifest as *chaotic regions* in the Poincaré section. From classical understanding of Hamiltonian chaos (Aref, 1984; Ottino, 1989), these chaotic dynamics arise as the stable and unstable manifolds no longer connect smoothly, resulting in a *chaotic saddle* within the open flow. This region is comprised of a fractal tangle of stable and unstable manifolds that imparts exponential stretching to fluid elements and a fractal distribution of residence times, see Tél et al. (2005), Toroczkai et al. (1998) for details. In contrast, elliptic points (E) represent regular, non-mixing regions in the flow. The Kolmogorov-Arnol'd-Moser (KAM) theorem states that the outermost orbits of these regular "islands" (termed KAM islands, see Figure 2d) around elliptic points are the least stable and so will break up into chaotic trajectories with further increases in the compressibility parameter $\mathcal{C}$ beyond $\mathcal{C}_2$. Eventually all of the orbits will become chaotic but the elliptic point itself will persist.

### 3.2.4.  Tidal Emptying Region and Summary

There also exists an inflow/outflow region near the tidal boundary termed the *tidal emptying region* (see TLMW for details), where fluid particles are swept in/out of the aquifer within one flow period $P$. This region is bounded by the tidal boundary ($x = 0$) on the left and the rightmost boundary of this region is termed as the *tidal emptying boundary*. The three transport structure types (open, closed, chaotic) shown in Figure 2 represent all of the different transport structures observed over the parameter space $\mathcal{Q} = \mathcal{T} \times \mathcal{G} \times \mathcal{C}$. In the following Section we then explore how these dynamics impact the global transport properties within the aquifer and the implications for solute transport.

## 4.  Parameter Sweep

To determine the range of transport dynamics and Lagrangian kinematics in periodically forced aquifers we compute the linear groundwater model over the aquifer parameter space $\mathcal{Q} = \mathcal{T} \times \mathcal{G} \times \mathcal{C}$. We analyse the transport structures outlined in the previous section in order to explore the distribution of Lagrangian kinematics over the parameter space and understand of how the underlying physical mechanisms represented by the dimensionless parameters control these transport structures. Unless stated otherwise, all simulations in this study use a $164 \times 164$ finite difference grid to resolve flow over the 2D hydraulic conductivity field with mean value $K_{\text{eff}} = 2 \times 10^{-4}$, log-variance ($\sigma^2_{\log K} = 2$) and correlation length ($\lambda = 0.049$). We use the finite-difference and particle mapping methods described in TLMW to provide numerical solutions to the flow field $\mathbf{v}(\mathbf{x}, t)$ and then generate the associated Poincaré sections over the flow parameter space $\mathcal{Q} = \mathcal{T} \times \mathcal{G} \times \mathcal{C} = (1, 10, 100) \times (1, 10, 100) \times (0, 0.01, 0.02, 0.05, 0.1, 0.2, 0.5)$ in an efficient manner. The rationale for selecting these parameter values is as follows. Whilst the tidal strength $\mathcal{G}$ can range from zero (no tidal flow) to infinity (no regional flow), non-trivial transport behaviour occurs for finite values of $\mathcal{G}$, hence we consider the intermediate values above. Similarly, the Townley number $\mathcal{T}$ ranges from infinity (no penetration of the tidal signal into the aquifer) to zero (infinite propagation of the tidal signal but of vanishing magnitude), so we consider intermediate values of $\mathcal{T}$ where complex transport arises. The compressibility parameter $\mathcal{C}$ ranges from zero (incompressible) to order unity (strongly compressible), but we note porosity values greater than 0.5 are rare and so consider the set of compressibility values above. In order to ensure robustness of the particle mapping method to solve (9), we tested two different integration schemes (explicit and implicit Runge-Kutta) and mapping grid resolutions ($200 \times 200$ and $400 \times 400$ grids over the aquifer domain), and found the results (Supplementary Information) to be insensitive to these changes. To test the impact of the complexity of the hydraulic conductivity field upon transport we also perform studies over the hydraulic conductivity parameter space $\chi$ spanned by the log-variance $\sigma^2_{\log K}$ and correlation length $\lambda$ as $\chi = (\sigma^2_{\log K}, \lambda) = (0.5, 1, 2, 4) \times (0.025, 0.049, 0.074, 0.098)$. We test for statistical invariance in all cases by performing simulations for three different realisations of the random hydraulic conductivity field. Of primary interest are the transport dynamics that arise within the periodically forced aquifer, the topological bifurcations between transport structures and the physical implications of the transport dynamics over the aquifer parameter space defined by $Q$ and $\chi$. This parametric sweep study elucidates potential ramifications for coastal groundwater systems, e.g. transport, mixing and chemical reactions and uncovers the physical mechanisms that control these processes.

## 4.1. Impact of tidal strength and attenuation on aquifer transport

To begin, we consider the impact of both tidal strength and attenuation on the transport dynamics in the tidally forced aquifer by varying the Townley number $\mathcal{T}$ and tidal strength $\mathcal{G}$ at a fixed value ($\mathcal{C} = 0.1$) of the tidal compression ratio. The Poincaré sections generated by the distributed injection protocol are shown in Figure 3, where the purple curves indicate the tidal emptying boundary, and the region between this boundary and tidal boundary ($x = 0$) is the tidal emptying region which indicates the part of the aquifer for which intense mixing between freshwater and saltwater occurs over a single flow period. Note also that the rightmost boundary (brown) in Figure 3 is the *inland boundary*, given by the location of particles that have been seeded along the inland boundary ($x = L$) after one flow period. Whilst this boundary gives some information regarding inland transport, this is of less physical importance than the tidal emptying boundary ($x = 0$) as the inland domain boundary ($x = L$) denotes the extent of the computational domain.

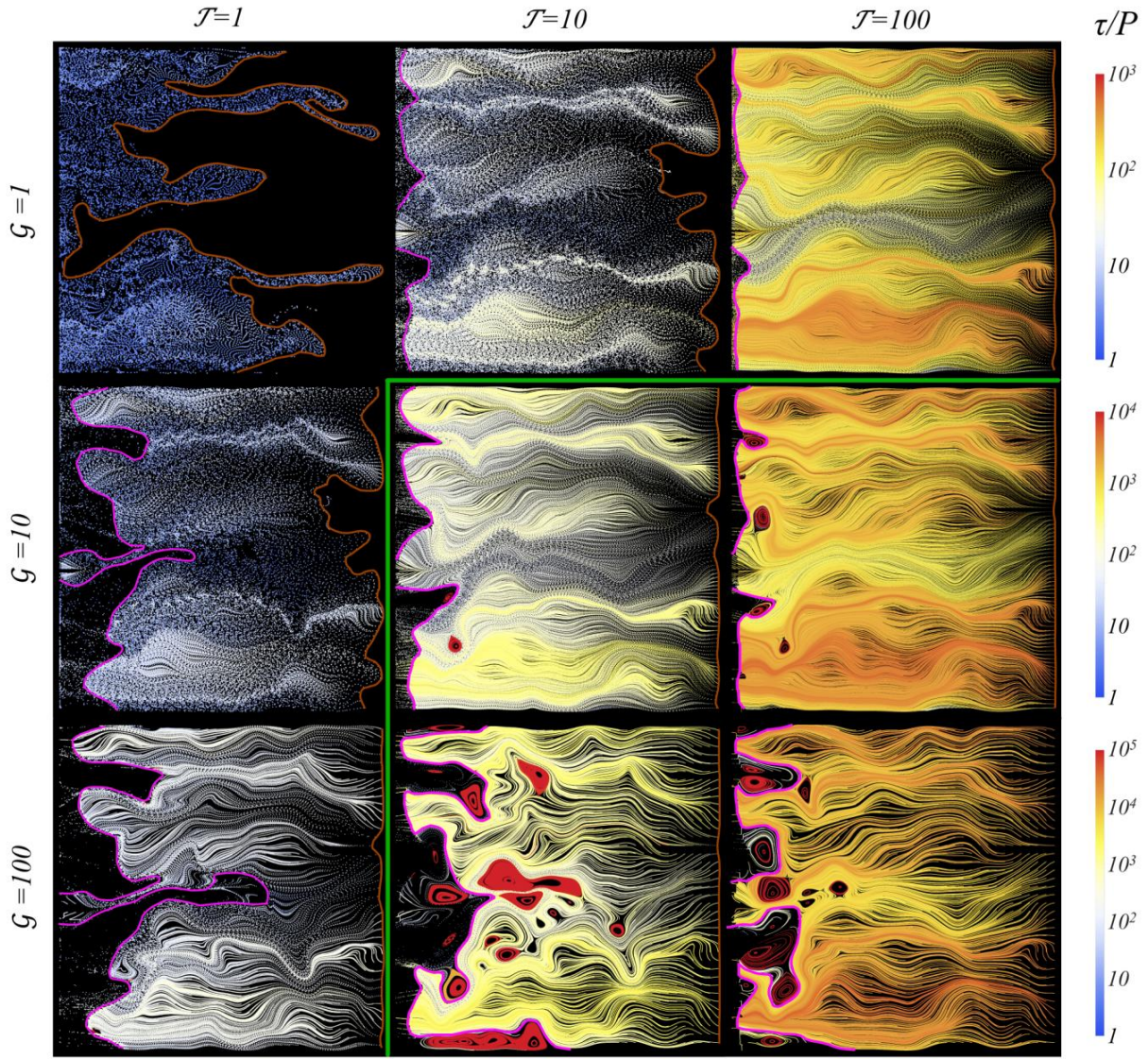

**Figure 3.** Summary of Poincaré sections for $\mathcal{C} = 0.1$ calculated via the distributed injection of particles. Particle trajectories are coloured according to residence time, from shortest (blue) to longest (red), and the colour scales change with $\mathcal{G}$. Green lines indicate Poincaré sections that possess trapped orbits. Purple and brown curves are the tidal emptying and inland boundaries respectively. The characteristics of the heterogeneous field: $K_{\text{eff}} = 2 \times 10^{-4}$, $\sigma^2_{\log \kappa} = 2$, $\lambda = 0.049$.

The grid of Poincaré sections shown in Figure 3 contains rich information about the transport dynamics throughout the aquifer domain and their variation with the control parameters. When the value of either the Townley number $\mathcal{T}$ or tidal strength $\mathcal{G}$ is small ($< 10$), all of the time-averaged particle orbits have open trajectories that travel from the inland boundary ($x = L$) to the tidal emptying boundary in a manner similar to that of steady Darcy flow. Conversely, when both $\mathcal{T}$ and $\mathcal{G}$ are larger ($> 10$), these time-averaged orbits undergo a topological bifurcation to admit localized closed orbits (trapping regions) and periodic points. This transition is predicted by the temporal character $\mathcal{H}_t$ which measures the propensity for flow reversal to occur and increases with both $\mathcal{T}$ and $\mathcal{G}$. If the value of $\mathcal{H}_t$ is large, particles will drift many tidal periods, but their elliptical velocity orbits become too small to generate significant flow reversal. The spatial character $\mathcal{H}_x$ provides information on how many vortices are likely to occur within the aquifer, so complex transport arises when $\mathcal{H}_x$ is greater than one. As shown in Figure 3, both $\mathcal{H}_x$ and the width of the tidally active zone decrease with increasing $\mathcal{T}$, and these values increase with increasing tidal strength $\mathcal{G}$. Hence, complex transport structures arise for moderate values of $\mathcal{T}$ and large values of $\mathcal{G}$. These trends persist for different values of the tidal compression ratio $\mathcal{C}$ beyond that ($\mathcal{C} = 0.1$) used in Figure 3.

The closed regions shown in Figure 3 occur in both the aquifer bulk and in the tidal emptying region, with significant implications for transport of solutes. If trapped regions in the tidal emptying region are initially filled with freshwater, then these trapped regions represent "islands" of freshwater within a "sea" of brackish water in the tidal emptying region, and salt may only enter via diffusion or dispersion. In Section 5.2 we show that these regions persist for long times and strong concentration gradients are expected at their boundaries. Conversely, trapped regions within the aquifer bulk may represent islands of different water composition to that of the fluid entering via the inland boundary ($x = L$), and the only transport mechanism into or out of these regions is diffusion or dispersion. Hence these augmented dynamics and concentration gradients will have strong impacts upon transport, mixing and chemical reactions within the aquifer.

As the parameter intervals for $\mathcal{T}$ and $\mathcal{G}$ in Figure 3 are too broad to accurately discriminate the flow structure transition boundary, we improve resolution of the parameter space to the values $\mathcal{T}, \mathcal{G} = (1, 2, 5, 10, 20, 50, 100)$ for $\mathcal{C} = 0.01$ and $\mathcal{C} = 0.1$ and summarise the observed transport structure types in Figure S9 of the supplementary material. Using this higher resolution, we observe a bifurcation to closed time-averaged orbits for a broad range of $\mathcal{T}$ values when $\mathcal{G} \gtrsim 50$ for $\mathcal{C} = 0.01$ and $\mathcal{G} \gtrsim 10$ for $\mathcal{C} = 0.1$. No evidence of chaotic trajectories was observed in these simulations for the tidal compression ratios $\mathcal{C} = 0.01$ and $0.1$. As expected, we observe an increase in the probability of closed orbits with increasing $\mathcal{T}, \mathcal{G}, \mathcal{C}$ across all cases computed.

### 4.2. Impact of aquifer compressibility on transport

To understand the role of aquifer compressibility on transport, we now vary the compressibility parameter $\mathcal{C}$. Figure 4 shows a series of aquifer Poincaré sections using the distributed injection protocol at various values of $\mathcal{C}$ in the range $\mathcal{C} = 0 - 0.5$ for $\mathcal{T} = 10$ and $\mathcal{G} = 100$. With increasing $\mathcal{C}$, the particle drift velocity increases and the tidal emptying region widens. In TLMW, we predicted that fluid tracer particle trajectories are integrable for incompressible ($\mathcal{C} = 0$) aquifers, which means they propagate in an oscillatory manner along smooth 1D streamlines and so cannot exhibit trapping or chaotic dynamics. Figure 4 confirms that particle trajectories are indeed confined to 1D orbits for $\mathcal{C} = 0$, and this behaviour appears to extend to very weakly compressible aquifers ($\mathcal{C} = 0 - 0.002$). Closed particle orbits are observed for weakly compressible aquifers ($\mathcal{C} = 0.005 - 0.1$), whereas strongly compressible aquifers ($\mathcal{C} = 0.2 - 0.5$) undergo a further topological bifurcation to exhibit chaotic transport structures, as indicated by the chaotic saddles highlighted in Figure 4.

As the region of parameter space ($\mathcal{T}$, $\mathcal{G}$, $\mathcal{C}$) for which chaotic transport is observed is a subset of that for which closed time-averaged orbits are observed, we hypothesise that as either $\mathcal{T}$, $\mathcal{G}$ or $\mathcal{C}$ are increased, the basic transport structure may bifurcate from open to closed orbits and admit periodic points that may be classed as elliptic (E) or hyperbolic (H). With further increases in $\mathcal{T}$, $\mathcal{G}$ or $\mathcal{C}$, the stable and unstable manifolds associated with the hyperbolic periodic points intersect transversely, corresponding to a bifurcation to chaotic transport dynamics, with important transport implications beyond that of simple trapping in closed orbits (see TLMW for details). Thus, there exists a hierarchy of transport complexity in periodically forced aquifers based on the route to chaos in aquifer flow structures as described above.

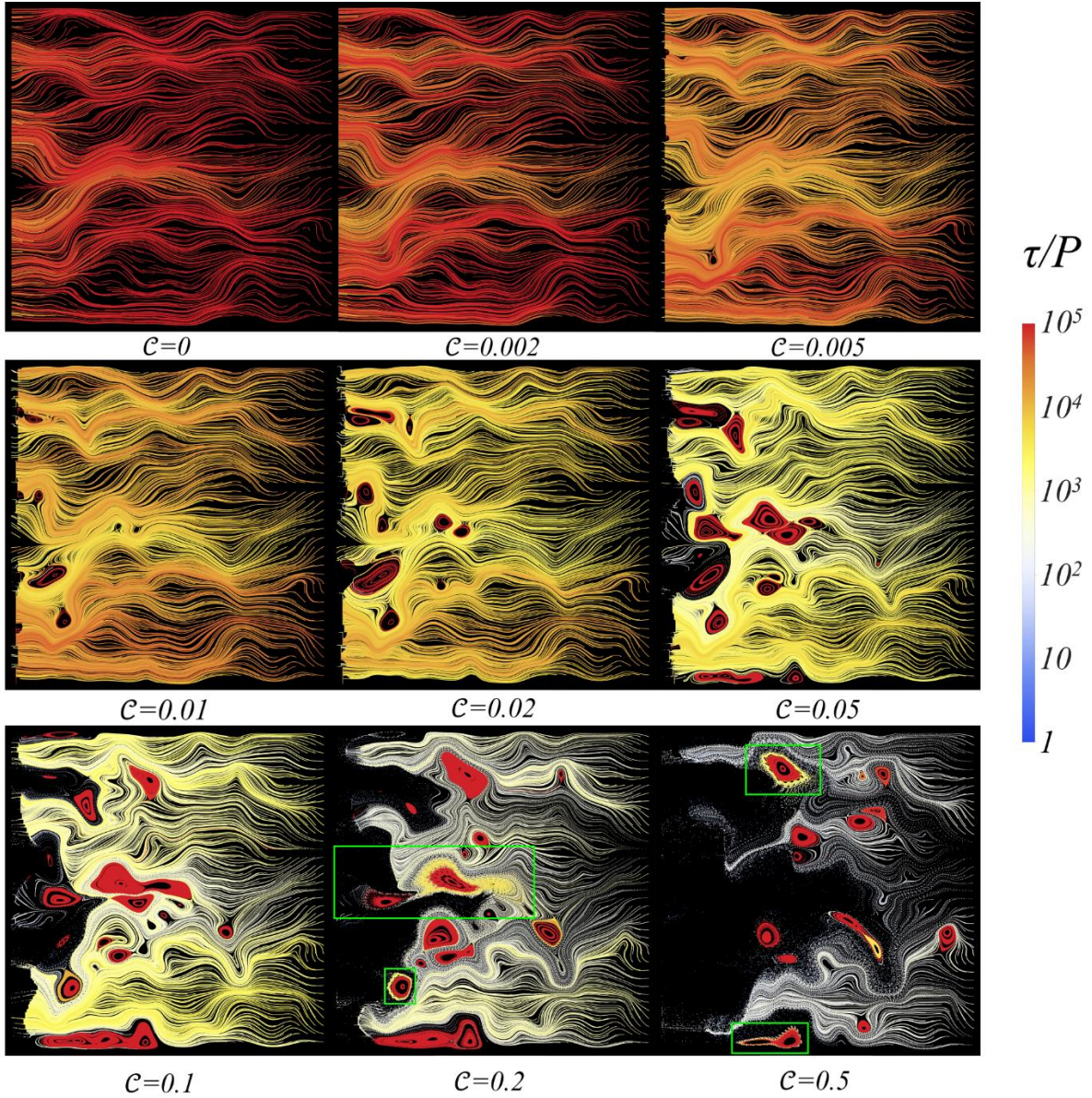

**Figure 4.** Poincaré sections in the periodically forced aquifer under the distributed injection protocol for different values of $\mathcal{C}$ over the range $\mathcal{C} = 0 - 0.5$ for $\mathcal{T} = 10$ and $\mathcal{G} = 100$. Green rectangles identify chaotic transport structures. The characteristics of the heterogeneous field: $K_{\text{eff}} = 2 \times 10^{-4}$, $\sigma^2_{\log K} = 2$, $\lambda = 0.049$.

### 4.3. Sensitivity to hydraulic conductivity field

All numerical treatments of Darcy flow in porous media depend on the model chosen to represent conductivity heterogeneity. Of primary interest, however, is the essential physics of the studied flow, i.e. those features which are robust to both the class and details of the conductivity model. In TLMW, we used a simple Gaussian autocorrelation function for log-conductivity that is statistically determined by the mean conductivity ($K_{\text{eff}}$), conductivity log-variance ($\sigma^2_{\log K}$) and correlation length ($\lambda$) to analyse the transport structure of the associated flow field. The concept is that if complex transport dynamics can occur in these simple conductivity fields, then it is possible and even likely that similar dynamics will occur in more sophisticated geo-statistical models. Here we examine how the transport structures of these flows vary with the parameters of the Gaussian autocorrelation model. Specifically, we consider the effects of different spatial realizations of the conductivity field, different correlation lengths ($\lambda = 0.025, 0.049, 0.074, 0.098$) and different log-variances ($\sigma^2_{\log K} = 0.5, 1, 2, 4$). Whilst these smaller correlation lengths and larger log-variances may demand greater numerical resolution to evaluate these flows, we find the original finite-difference grid is sufficient to accurately resolve these cases. We also study sensitivity of the type and number of predicted transport structures by generating a set of different realizations of the log-conductivity fields according to the algorithm of Ruan and McLaughlin (1998).

#### 4.3.1. Impact of conductivity variance

We measure the effect of different log-conductivity variances ($\sigma^2_{\log K} = 0.5, 1, 2, 4$) with fixed correlation lengths $\lambda = 0.049$. As shown in TLMW, flow reversal plays a critical role in generating complex transport dynamics. Log-conductivity variance controls the magnitude of vorticity and so is strongly correlated with flow reversal. Hence, the density of canonical flux ellipses is related to the reversal number $\mathcal{G}\,\sigma^2_{\log K}$. Figure 5 shows that as the reversal number increases either through increase of $\mathcal{G}$ or increase of $\sigma^2_{\log K}$, the number and size of closed flow regions increases until these closed regions bifurcate into chaotic regions (for $\sigma^2_{\log K} = 4$), and in the supplementary material (S3) we show that this trend persists for different values of $\mathcal{T}$ and $\mathcal{C}$.

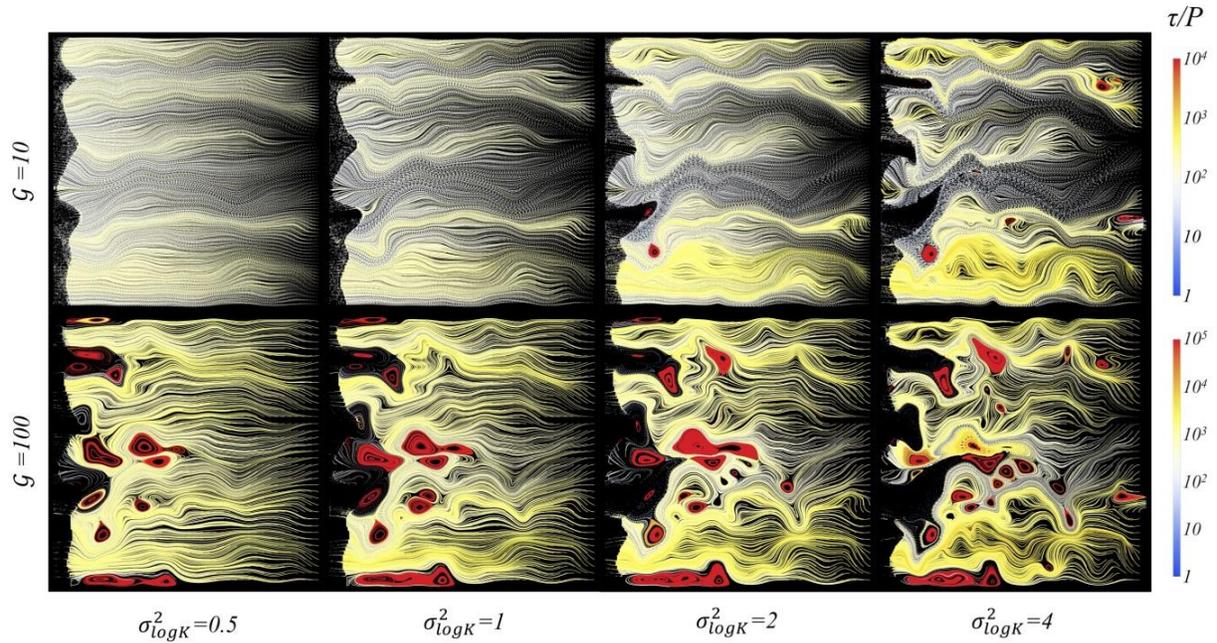

**Figure 5.** Poincaré sections (distributed injection) of flow in a single log-$K$ realization for $\mathcal{T} = 10$ and $\mathcal{C} = 0.1$ with varying $\mathcal{G}$ and $\sigma^2_{\log K}$. The characteristics of the heterogeneous field: $K_{\mathrm{eff}} = 2\times10^{-4}$, $\lambda = 0.049$.

In the supplementary material (S2) we show that the model predictions are statistically invariant with respect to different realizations of the random conductivity field. In S4 of the supplementary material we also show that the correlation length is a key aquifer property that controls the size of closed regions within the flow. These results indicate that the Lagrangian complexity of the periodic flow system varies smoothly with key heterogeneity parameters (correlation length, log-variance) of the Gaussian statistical model, and that our observations of transport dynamics are robust to common heterogeneity variations. Further studies using different statistical models for the heterogeneity field, such as multi-Gaussian, kriged and fractal models, would provide valuable checks and further insight.

## 5. Solute trapping and transport

In this Section we consider the impact of closed transport regions upon solute transport by considering both the size distribution of closed regions across the parameter space, and the impact of the size of trapped regions upon solute diffusion and dispersion.

### 5.1. Size distribution of closed regions

When closed regions do occur within the aquifer, both the location (i.e. whether in the tidal emptying region or the aquifer bulk) and size distribution of these regions have important implications for solute transport, as solutes will more rapidly disperse out of smaller closed regions. We use tracer particle RTDs under the distributed injection protocol to determine the total of the area ($a$) of trapped regions in the aquifer. The resultant RTD histogram (probability density function) for $\mathcal{T} = 100$, $\mathcal{G} = 100$ and $\mathcal{C} = 0.1$ is shown in Figure S10 of the supplementary material. Particle tracking is performed up to $\tau/P = 10^5$ forcing periods, and most of the tracer particles leave the domain within $4{\times}10^4$ forcing periods. A significant proportion (4.28%) of tracer particles are confined indefinitely which are deemed to be trapped regions.

The inset of Figure S10a summarises the total of the relative area (*a*) of trapped regions (closed orbits) in the aquifer for various values of $\mathcal{T}$, $\mathcal{G}$ at $\mathcal{C} = 0.1$. Whilst the total trapped area is large for $\mathcal{T} = 10$, $\mathcal{G} = 100$, the total trapped area is smaller for $\mathcal{T} = 100$, $\mathcal{G} = 100$ due to the smaller tidal region shown in Figure 3. Figure S10b indicates that trapped regions are typically smaller for small values of $\mathcal{T}$ and $\mathcal{G}$, but with parameter increases the trapped regions increase in size and coalesce into fewer but larger regions. As we show explicitly in the next section, larger trapped regions trap diffusive solutes for much longer time and so represent significant regions of retarded transport.

## 5.2. Solute transport across trapped regions

Whilst the analysis in this study focuses predominantly upon the advection processes that govern these transport structures, solutes are typically transported by advection and diffusion/ dispersion. In the presence of dispersive transport mechanisms, closed regions of the flow become regions of retarded transport. To provide and order-of-magnitude estimate of the transport rate of solutes into or out of these regions, we approximate the trapped region as a disc of radius $R_0 L$ (such that $R_0$ is the disc radius relative to the aquifer dimension $L$) that is surrounded by fluid with zero solute concentration in the open region of the flow. This estimate considers the dispersive "leakage" of the solute concentration field $\mathbf{c}(\mathbf{x}, t)$ out of such a trapped region subject to homogeneous Dirichlet conditions (c = 0) at the boundary. Under these assumptions, the solute concentration field is governed by the dispersion equation

$$\frac{\partial c}{\partial t} = \nabla \cdot (\mathbf{D}_H \cdot \nabla c),$$ (11)

where $\mathbf{D}_H$ is the hydrodynamic dispersion tensor.

### 5.2.1. Diffusive transport

We first consider the case of molecular diffusion, in which case $\mathbf{D}_H = D_m \mathbf{I}$, where $D_m$ is the molecular diffusivity. In this case the total amount of solute $C(t) \in [0,1]$ in the trapped region relative to the initial amount (such that $C(0) = 1$) is

$$C(t) = \sum_{n=1}^{\infty} \frac{4}{\alpha_n^2} e^{-\frac{1}{\text{Pe}_m} \frac{\alpha_n^2 t}{R_0^2 P}} \approx \frac{4}{\alpha_1^2} e^{-\frac{1}{\text{Pe}_m} \frac{\alpha_n^2 t}{R_0^2 P}}, \tag{12}$$

where $\alpha_n$ is the $n$-th zero of the zero-th order Bessel function of the first kind and $\text{Pe}_m = L^2/(D_m P)$ is a Péclet number defined in terms of the forcing period $P$ and dimension $L$ of the aquifer. The molecular diffusivity of water $D_m \sim 10^{-9}$ m²/s then yields an estimate of this Péclet number of order $\text{Pe}_m \sim 10^{11}$ for a typical coastal aquifer of width $L \sim 2000$m subject to a $P = 12$ hour tidal period. As per equation (12), for all but very short times, the total relative solute amount decays exponentially at a rate very close to $-\alpha_1^2/(R_0^2 \text{Pe}_m)$ (with $\alpha_1 = 2.40483$), thus 90% of solute has leached out of the trapped regions after time $T_{90\%}$, which is defined as

$$T_{90\%} \approx P \, \text{Pe}_m \frac{R_0^2}{\alpha_1^2} \ln(10). \tag{13}$$

Thus, the residence time of a solute within a given trapped region scales as $R_0^2$, which is proportional to the trapped region area ($a$). Hence the histograms of the trapped area size ($a$) shown in **Error! Reference source not found.**b can be used to directly inform how solute transport varies with the aquifer parameters $\mathcal{T}$, $\mathcal{G}$, $\mathcal{C}$. In the supplementary material (S5) we also show that the relative radius $R_0$ of trapped regions scales linearly with the correlation length scale at a rate that varies with the Townley number $\mathcal{T}$, hence the duration of solute trapping scales quadratically with the correlation length scale. Figure S10**Error! Reference source not found.**b shows that a typical closed region has relative area of $a \sim 0.001$, and so the corresponding disc has relative radius $R_0 \sim$ 0.0178 and from (13) this takes around $T_{90\%} \sim 16{,}000$ years to empty. Hence these trapped regions have a profound impact upon transport of diffusive solutes.

### 5.2.2. Dispersive transport

To provide a lower bound for the emptying time for solute transport under hydrodynamic dispersion we replace the hydrodynamic dispersivity with $\mathbf{D}_H \rightarrow D_L \mathbf{I}$ in (13) and use the Scheidegger parameterization for the longitudinal $D_L$ component (which is significantly larger than the transverse component) of the hydrodynamic dispersivity tensor which is related to the local velocity as

$$D_L = D_m + \alpha_L v, \qquad (14)$$

where the longitudinal dispersivity for a coastal aquifer $\alpha_L$ is estimated to be of order 10 m or larger. In the supplementary material (S6) we show that the regional flow velocity $v_r = J\, K_{\text{eff}}/\varphi_{\text{eff}}$ provides a reasonable estimate of the local velocity $v$ in (14), and for typical values $J = 0.005$, $K_{\text{eff}} = 10$ m/day, $\varphi_{\text{eff}} = 0.3$ yields $v_r \sim 0.16667$ m/day, which gives an estimate of the longitudinal dispersivity of $D_L \sim 2 \times 10^{-5}$ m$^2$/s, and a corresponding Peclet number of Pe $\sim 10^6$. Equation (13) then predicts a solute emptying time of 304 days, suggesting trapped regions can still dominate solute transport in the presence of hydrodynamic dispersion.

## 6. Physical relevance

The parametric studies in the previous section clearly show how complex transport is a natural consequence of matrix heterogeneity and compressibility in formations subject to periodic forcing. Nevertheless, it may not yet be clear when and where to expect complex transport in natural aquifers. TLMW showed that physically plausible values for $\mathcal{T}$ and $\mathcal{G}$ yielded Poincaré sections exhibiting elliptic and hyperbolic points at high tidal compressibility $\mathcal{C} = 0.5$, and results from the present work have extended the Lagrangian complexity regime to much lower tidal compressibility values. In this section we seek to (i) clarify how realistic coastal aquifer systems relate to the topological bifurcation phenomena and (ii) provide the reader with a simple rule of thumb for assessing the potential for Lagrangian complexity in any particular field setting. In this analysis, we maintain the single-mode spectral approximation for the tidal forcing, although it is recognized that tidal spectra commonly display multimodal forms, and the effects of multimodal forcing is the subject of future studies.

## 6.1. Tidal transition diagram

As discussed in TLMW, published coastal aquifer studies can be used to assess likely parameter ranges for common coastal aquifer types including sediments, limestones, fractured basalts, etc. However, the Lagrangian flow complexity is also coupled with the effective coastal tidal amplitude which is a complicated function of position around the globe. We can decouple this complexity by plotting a small number of published field sites in $\mathcal{T} \times \mathcal{G} \times \mathcal{C}$ parameter space, superimposing the results of the topological analysis, and seeking zones of overlap for increasing tidal amplitude. We term such a plot the *tidal transition diagram* as it maps how tidal flows in specific aquifer formations move toward Lagrangian complexity as the effective tidal amplitude increases. The tidal transition diagram in Figure 6 is constructed as follows:

i.  A rectangular prism is constructed in $\mathcal{T} \times \sigma^2 \mathcal{G} \times \mathcal{C}$ in log-scale space.

ii.    Three-dimensional coloured ellipsoids, with ellipse axes approximating error bars on the parameter estimates, are plotted according to the parameters reported in the published field studies discussed in TLMW.

iii.   Shaded zones are superimposed to indicate the parametric extents of the elliptic bifurcation zone (pink) and the zone of chaotic flow (red) found in the present simulations.

iv.    Phase lines are plotted from the centroid of each ellipsoid to chart how its position would change for varying tidal amplitude $g_p$ in the range [0.01, 8] where the upper bound is the maximum tidal amplitude observed on earth (Bay of Fundy, 8 metres).

The phase lines can be computed as exact linear functions of $g_p$ from the definitions for $\mathcal{G}$ and $\mathcal{C}$ in Section 2. Figure 6 shows that the tidal trajectories are increasing functions of $g_p$ and display slightly curvilinear profiles since the $\mathcal{C}$ axis is not drawn to be exactly logarithmic. Notably, all aquifer types impinge on the pink elliptic bifurcation zone as $g_p$ increases, at amplitudes as low as 2m, see notation marks in Figure 6b. Chaotic phenomena (red zone) are rarer, at least for this small set of field sites. Observed tidal amplitudes are complicated functions of geographic location, however Figure 7 shows significant stretches of coastline where either K1 diurnal or M2 semidiurnal modes dominate, and in other areas mixed tides (with both diurnal and semidiurnal modes) add richness to the tidal forcing signal. There are extensive zones around the world where tidal amplitudes are much larger than for the limited case studies used here and in TLMW. Thus, we may anticipate that the pre-requisites for complex Lagrangian flows (compressible coastal aquifers and large tidal amplitudes) are not uncommon at a global scale. Ideally, we would also like to map aquifer properties (compressibility, heterogeneity etc) and the strength of regional flow to conclusively identify regions where complex transport is likely, however this information is only know at a handful of sites around the globe, some of which were considered in TLMW.

These results indicate that the interplay of periodic forcing and regional flow in coastal aquifer systems can generate complex transport dynamics similar to those observed (Cho et al., 2019; Libera et al., 2017; Metcalfe et al., 2010a; 2010b; Trefry et al., 2012) in transient pumping schemes. Steady and transient pumping schemes are also used in some coastal aquifers to retard saltwater intrusion, and so an important practical question is how such pumping might impact the transport mechanisms discussed herein. Whilst such questions are beyond the scope of this study, we note that the tools and techniques utilized in this study may be readily extended to answer such questions and identify optimal pumping protocols to control solute transport and minimise intrusion.

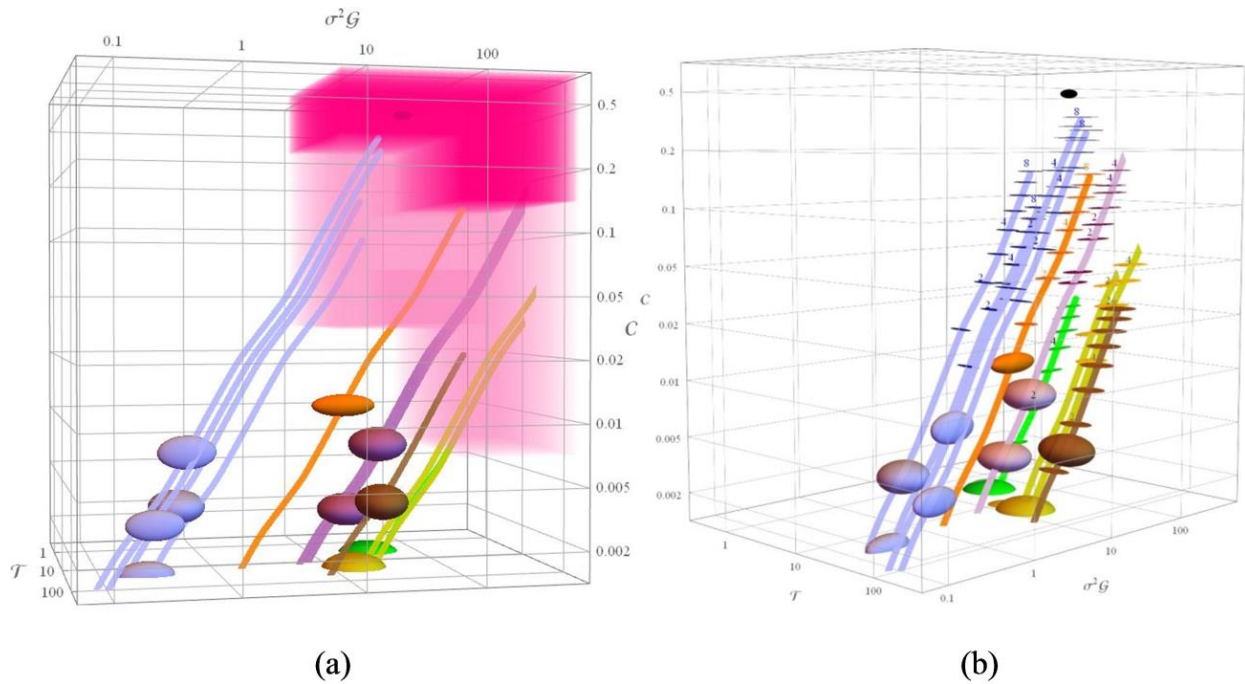

**Figure 6.** Tidal transition diagram showing (a) bifurcation and chaotic zones, and (b) tidal amplitude ($g_p$) trajectories marked each metre up to a maximum of 8 m. Ellipsoids refer to the GI (blue, limestone and sand), LN (orange, sand and clay), J (purple, limestone and sand), D (brown, limestone and sand), SR (yellow, sand and clay) and P (green, basalts) case studies of TLMW and the small black ellipse marks the example problem $(\mathcal{T}, \sigma^2\mathcal{G}, \mathcal{C}) = (10\pi, 20, 0.5)$. Trajectories reach

the elliptic zone (pink) at values g$_p$ = 3 m (GI), 1.5 m (LN), 0.5 m (J), 3.5 m (D), 0.5 m (SR), 2 m (P).

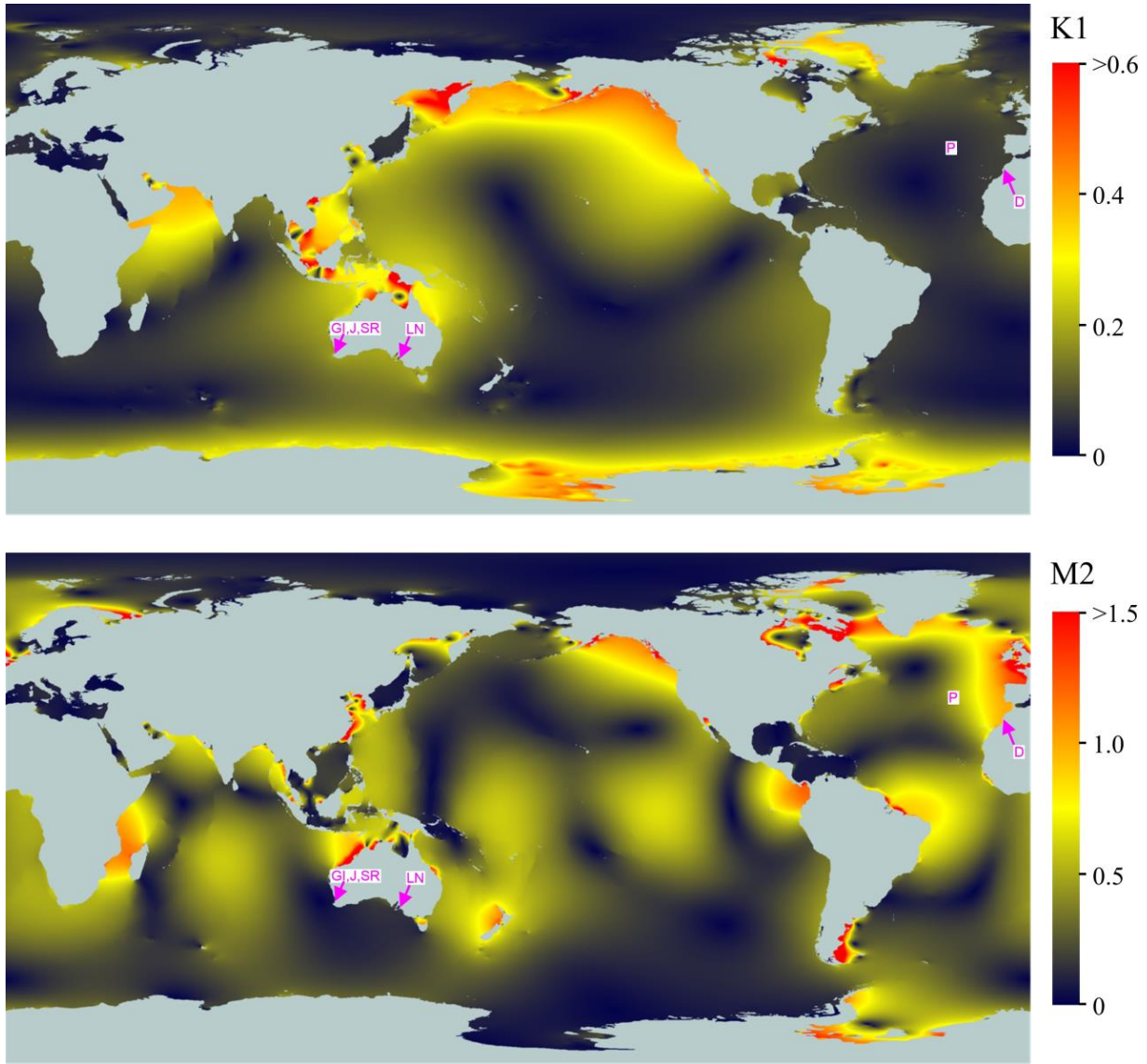

**Figure 7.** Global variation of the amplitudes (in metres) of the K1 (diurnal, top) and M2 (semi-diurnal, bottom) tidal modes (FES2004 data supplied by LEGOS, see Lyard et al. (2006)). Site locations used in Figure 6 are indicated by pink lettering. Zones of high amplitude indicate increased potential for Lagrangian complexity in coastal groundwater flows.

## 7. Conclusions

TLMW first demonstrated that the interplay of tidal forcing, aquifer heterogeneity and poroelasticity can produce complex Lagrangian transport, which has profound ramifications for solute transport and reaction in transiently forced groundwater systems. What was unknown, however is to what extent these transport dynamics arise in natural groundwater systems and how these dynamics depend upon the aquifer properties and characteristics of the regional and tidal flows. In this study we answered these questions through parametric studies of the governing parameters that characterize the aquifer and tidal properties. This study elucidates the mechanisms that govern complex transport dynamics, bifurcations between structures with changes in the governing parameters, and indicates the propensity of complex dynamics to occur in natural aquifer systems.

To achieve this, we defined two distinct parameter sets; the aquifer parameter set $\mathcal{Q} = (\mathcal{T}, \mathcal{G}, \mathcal{C})$ which controls the dynamics of a periodically forced aquifer, and the hydraulic conductivity parameter set $\chi = (K_{\text{eff}}, \sigma_{\log K}^2, \lambda)$ which characterises the log-Gaussian conductivity field. We used numerical simulation to consider the transport dynamics of a model tidally forced aquifer over these two parameter sets. We observed a change from open fluid particle orbits (similar to open streamlines characteristic of steady Darcy flow) to closed orbits that can trap fluid particles for indefinitely long periods. With further parametric variation, these closed orbits can then become chaotic, leading to accelerated mixing dynamics and slow "leaking" of particles into and out of these previously closed regions. We find that these three different transport structure types (open, closed, chaotic) arise at particular combinations of the governing parameters, leading to a "phase portrait" of transport dynamics over the aquifer parameter space that we explain via the physical mechanisms that govern flow in periodically forced aquifers.

Stability and robustness of the Gaussian statistical model have also been tested via changing the correlation length ($\lambda$), log-variance ($\sigma^2_{\log K}$) and log-conductivity realization, and we find that the size of closed regions increases linearly with the correlation length ($\lambda$), and the propensity for complex transport increases with log-variance ($\sigma^2_{\log K}$). Although we chose a simple log-conductivity field in this study, these results give confidence that the dynamical parameters $Q$ will likely generate similar flow characteristics regardless of the aquifer random heterogeneity models employed.

We have quantified the impact of closed and chaotic regions upon transport of diffusive and dispersive solutes and find that such trapping impacts solute transport over long timescales at a rate proportional to the size of the closed flow region. We also examine parametric variation of the size distribution of trapped regions with the aquifer, allowing the impacts on solute transport to be directly estimated. Simple calculations show that common coastal aquifer types can display Lagrangian bifurcations to trapping flows for sufficiently high tidal amplitudes; such tidal amplitudes may be prevalent around the world, e.g. in northern Europe, east Africa and the Alaskan-Canadian Pacific coast. These phenomena may profoundly impact biogeochemical processes in coastal aquifers.

The results in this paper predict that complex transport structures and dynamics can arise in natural tidally forced aquifers around the world. When such dynamics do arise, they will lead to mixing and reactive transport patterns that are very different to those predicted by conventional Darcy flow models.

# Acknowledgements

Data employed in this work can be obtained from the cited references and from the supporting information.

# References

Aref, H. (1984). Stirring by chaotic advection. *Journal of Fluid Mechanics, 143*, 1-21. doi: https://doi.org/10.1017/S0022112084001233

Bear, J. (1972). *Dynamics of Fluids In Porous Media*. New York: Dover: American Elsevier Publishing Company.

Cho, M. S., Solano, F., Thomson, N. R., Trefry, M. G., Lester, D. R., & Metcalfe, G. (2019). Field Trials of Chaotic Advection to Enhance Reagent Delivery. *Ground Water Monitoring and Remediation, 39*(3), 23-39. doi:https://doi.org/10.1111/gwmr.12339

Coussy, O. (2004). *Poromechanics*: Wiley.

Dentz, M., Lester, D. R., Le Borgne, T., & de Barros, F. P. J. (2016). Coupled continuous-time random walks for fluid stretching in two-dimensional heterogeneous media. *Physical Review E, 94*(6), 061102. doi: https://doi.org/10.1103/PhysRevE.94.061102

Heiss, J. W., Post, V. E. A., Laattoe, T., Russoniello, C. J., & Michael, H. A. (2017). Physical Controls on Biogeochemical Processes in Intertidal Zones of Beach Aquifers. *Water Resources Research, 53*(11), 9225-9244. doi: https://doi.org/10.1002/2017wr021110

Lester, D. R., Metcalfe, G., Trefry, M. G., Ord, A., Hobbs, B., & Rudman, M. (2009). Lagrangian topology of a periodically reoriented potential flow: Symmetry, optimization, and mixing. *Physical Review E, 80*(3), 036208. doi: https://doi.org/10.1103/PhysRevE.80.036208

Lester, D. R., Rudman, M., Metcalfe, G., Trefry, M. G., Ord, A., & Hobbs, B. (2010). Scalar dispersion in a periodically reoriented potential flow: Acceleration via Lagrangian chaos. *Physical Review E, 81*(4), 046319. doi: https://doi.org/10.1103/PhysRevE.81.046319

Li, L., Barry, D. A., Jeng, D.-S., & Prommer, H. (2004). Tidal dynamics of groundwater flow and contaminant transport in coastal aquifers. *Coastal aquifer management: Monitoring, modeling, and case studies*, 115-141.

Libera, A., de Barros, F. P. J., & Guadagnini, A. (2017). Influence of pumping operational schedule on solute concentrations at a well in randomly heterogeneous aquifers. *Journal of Hydrology, 546*, 490-502. doi:https://doi.org/10.1016/j.jhydrol.2016.12.022

Liu, Y., Jiao, J. J., & Liang, W. (2018). Tidal Fluctuation Influenced Physicochemical Parameter Dynamics in Coastal Groundwater Mixing Zone. *Estuaries and Coasts, 41*(4), 988-1001. doi: https://doi.org/10.1007/s12237-017-0335-x

Lyard, F., Lefevre, F., Letellier, T., & Francis, O. (2006). Modelling the global ocean tides: modern insights from FES2004. *Ocean dynamics, 56*(5-6), 394-415. doi:https://doi.org/10.1007/s10236-006-0086-x

Metcalfe, G., Lester, D., Ord, A., Kulkarni, P., Rudman, M., Trefry, M., . . . Morris, J. (2010a). An experimental and theoretical study of the mixing characteristics of a periodically reoriented irrotational flow. *Philosophical Transactions of the Royal Society A: Mathematical, Physical and Engineering Sciences, 368*(1918), 2147-2162. doi:https://doi.org/10.1098/rsta.2010.0037

Metcalfe, G., Lester, D., Ord, A., Kulkarni, P., Trefry, M., Hobbs, B. E., . . . Morris, J. (2010b). A partially open porous media flow with chaotic advection: towards a model of coupled fields. *Philosophical Transactions of the Royal Society A: Mathematical, Physical and Engineering Sciences, 368*(1910), 217-230. doi:https://doi.org/10.1098/rsta.2009.0198

Ottino, J. M. (1989). *The Kinematics of Mixing: Stretching, Chaos, and Transport*. Cambridge, UK: Cambridge University Press.

Pool, M., Carrera, J., Dentz, M., Hidalgo, J. J., & Abarca, E. (2011). Vertical average for modeling seawater intrusion. *Water Resources Research, 47*(11). doi:https://doi.org/10.1029/2011wr010447

Pool, M., Post, V. E. A., & Simmons, C. T. (2015). Effects of tidal fluctuations and spatial heterogeneity on mixing and spreading in spatially heterogeneous coastal aquifers. *Water Resources Research, 51*(3), 1570-1585. doi:https://doi.org/10.1002/2014wr016068

Robinson, C., Brovelli, A., Barry, D. A., & Li, L. (2009). Tidal influence on BTEX biodegradation in sandy coastal aquifers. *Advances in water resources, 32*(1), 16-28. doi: https://doi.org/10.1016/j.advwatres.2008.09.008

Ruan, F., & McLaughlin, D. (1998). An efficient multivariate random field generator using the fast Fourier transform. *Advances in water resources, 21*(5), 385-399. doi:https://doi.org/10.1016/S0309-1708(96)00064-4

Slomp, C. P., & Van Cappellen, P. (2004). Nutrient inputs to the coastal ocean through submarine groundwater discharge: controls and potential impact. *Journal of Hydrology, 295*(1-4), 64-86. doi: https://doi.org/10.1016/j.jhydrol.2004.02.018

Tél, T., de Moura, A., Grebogi, C., & Károlyi, G. (2005). Chemical and biological activity in open flows: A dynamical system approach. *Physics Reports, 413*(2), 91-196. doi:https://doi.org/10.1016/j.physrep.2005.01.005

Toroczkai, Z., Károlyi, G., Péntek, Á., Tél, T., & Grebogi, C. (1998). Advection of Active Particles in Open Chaotic Flows. *Physical review letters, 80*(3), 500-503. doi: https://doi.org/10.1103/PhysRevLett.80.500

Trefry, M. G., Lester, D. R., Metcalfe, G., Ord, A., & Regenauer-Lieb, K. (2012). Toward enhanced subsurface intervention methods using chaotic advection. *J Contam Hydrol, 127*(1-4), 15-29. doi: https://doi.org/10.1016/j.jconhyd.2011.04.006

Trefry, M. G., Lester, D. R., Metcalfe, G., & Wu, J. (2019). Temporal Fluctuations and Poroelasticity Can Generate Chaotic Advection in Natural Groundwater Systems. *Water Resources Research, 55*(4), 3347-3374. doi:https://doi.org/10.1029/2018wr023864

Supporting Information for

**When do complex transport dynamics arise in natural groundwater systems?**

J. Wu[1], D. R. Lester[1], M. G. Trefry[2], G. Metcalfe[3],

[1] School of Engineering, RMIT, Melbourne, Victoria, Australia.

[2] Independent Researcher, Floreat, Western Australia, Australia.

[3] School of Engineering, Swinburne University of Technology, Hawthorn, Victoria, Australia.

**Contents of this file**

**Introduction**

Text S1 describes the mapping method for rapid solution of the advection equation and Figure S2 illustrates the impact of altering the grid resolution for this method.

Text S2 and Figure S2 illustrate the impact of arbitrary realizations for the log-conductivity field.

Text S3 and Figure S3 and S4 illustrate the impact of log-variance upon transport structures for different values of the Townley number and compressibility parameter.

Text S4 and Figures S5 and S6 illustrate the effect of different correlation lengths upon the aquifer transport structures.

Text S5 and Figure S7 shows the distribution of $R_0^2$ and $R_0^2/v$ for various values of the Townley number and tidal strength.

Text S6 provides an estimate of the local velocity of fluid elements in periodically forced aquifers.

Text S7 provides an estimate of the penetration distance of saline wedges in coastal aquifers relative to the penetration distance of tidal signals.

Text S8 summarises the propensity for closed regions to occur with variations in $\mathcal{T}$ and $\mathcal{G}$.

Text S9 illustrates how the residence time distribution and size of trapped regions vary with $\mathcal{T}$ and $\mathcal{G}$.

### Text S1. Mapping Method

We employ a "mapping method" to rapidly advect large numbers of fluid particles based upon solution of the advection equation (13) in the manuscript. This method involves advecting a $N \times N$ uniform square grid (termed an "integration grid") of tracer particles (where $N$ is typically 200 or greater) that covers the entire computational domain (x, y) = [0,1] x [0,1] over a single forcing period using a fourth order explicit Runge-Kutta method to solve the advection equation. This method was also tested using an implicit Runge-Kutta method and a finer ($N = 400$) grid, and no significant changes to the predicted transport dynamics were observed (see Figure S1). Any particles that leave the fluid domain are deemed to lie on the boundary location at which they leave the domain during the forcing cycle. The updated particle locations are then used to construct the following spline interpolation functions for the *x*- and *y*-coordinates at the end of a forcing period ($x_{n+1}, y_{n+1}$) as a function of the coordinates $(x_n, y_n)$ at the start of the forcing period:

$$x_{n+1} = f_x(x_n, y_n), \qquad y_{n+1} = f_y(x_n, y_n). \qquad (S1)$$

In TLMW we showed that the fluid velocity is net divergence-free over a single forcing period of the aquifer, hence in principle, this method should yield an area-preserving mapping such that the Jacobian $J = \det(\nabla f)$, $f = (f_x, f_y)$ is everywhere unity. However, interpolation of particle trajectories via equation (S1) above introduces small (typically $10^{-3}$) deviations in the Jacobian from unity. Whilst these deviations can produce spurious behaviour in e.g. chaotic regions of the flow over very large particle residence times, these effects are relatively minor and do not qualitatively impact the flow topologies in these regions. As expected, large deviations in the Jacobian also occur near the inflow/outflow domain boundaries due to the trapping of particles here, however these errors are highly localised and do not impact the internal transport structures. Once the mapping functions $(f_x, f_y)$ are generated, these may be used to rapidly advect large numbers of particles of many flow periods, facilitating rapid generation of the Poincare sections used in this study.

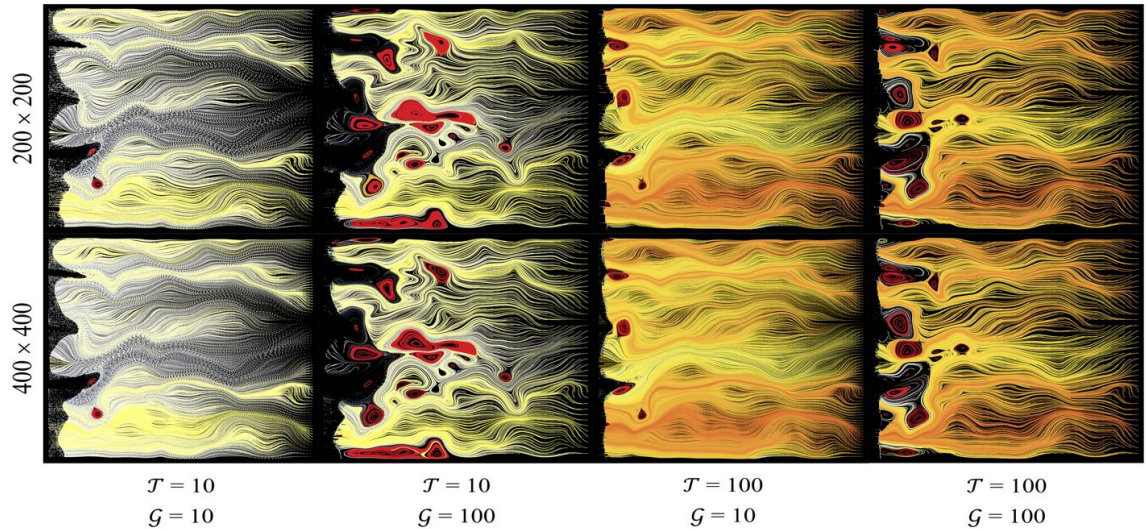

**Figure S1**. Impact of resolution of integration grid resolution $N$ (top row $N = 200$, bottom row $N = 400$) upon Poincaré sections for various values of the Townley number $\mathcal{T}$, tidal ratio $\mathcal{G} = 10$. The characteristics of the heterogeneous field: $K_{\text{eff}} = 2 \times 10^{-4}$, $\sigma_{\log \kappa}^2 = 2$, $\lambda = 0.049$.

**Text S2. Impact of conductivity realization**

Figure S2 shows three arbitrary realizations of the log-conductivity field (for fixed $K_{\text{eff}}$, $\sigma_{\log K}^2$ and $\lambda$), with the corresponding shaded density maps in the top row. For each realization, the Poincaré sections are provided for two different dynamical parameter sets (see figure caption for details). The three realizations provide similar flow topologies for each of the two dynamical parameter sets, i.e. within each row the numbers and densities of periodic points are similar, as are the spatial extents of the tidal emptying zones. Further testing indicates that these similarities persist between realizations over a broader range of dynamical parameters (results not shown). Thus, there is no essential change in the Lagrangian topologies arising from change of log-conductivity realization, although the Poincaré sections differ in detail.

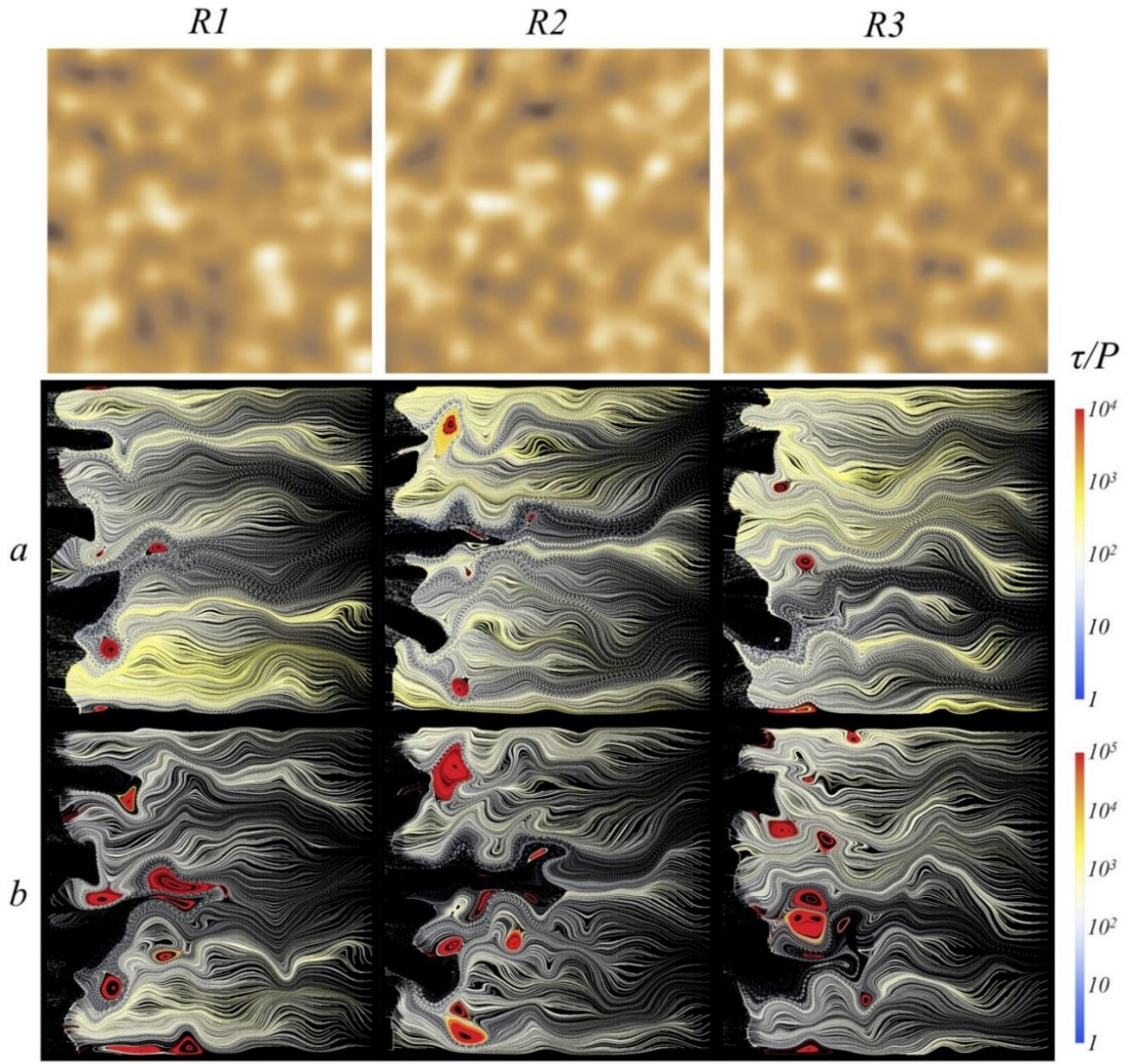

**Figure S2**. Density map of log hydraulic conductivity for three different realizations of the log-Gaussian hydraulic conductivity field and associated Poincaré sections (distributed injection) for two cases, row *a*: $\mathcal{T}$ = 20, $\mathcal{G}$ = 10, $\mathcal{C}$ = 0.2, row *b*: $\mathcal{T}$ = 5, $\mathcal{G}$ = 50, $\mathcal{C}$ = 0.1. The characteristics of the heterogeneous field: $K_{\text{eff}}$ = 2×10$^{-4}$ , $\sigma_{\log K}^2$ = 2, $\lambda$ = 0.049.

**Text S3. Impact of conductivity log-variance for various values of $\mathcal{C}$ and $\mathcal{T}$**

In the main paper, we consider the impact of varying log-variance of the conductivity field for the dimensionless parameters $\mathcal{T} = 10$, $\mathcal{G} = 10$ and 100, and $\mathcal{C} = 0.1$. In Figure S3 and S4 respectively we also show the impact of conductivity variance for different values of $\mathcal{C}$ and $\mathcal{T}$ for $\mathcal{G}$ =10 and all other parameters as above unless stated otherwise. As for the cases shown in the main paper, the complexity of transport dynamics increases with conductivity log-variance, including the number of closed regions and the propensity for chaotic mixing.

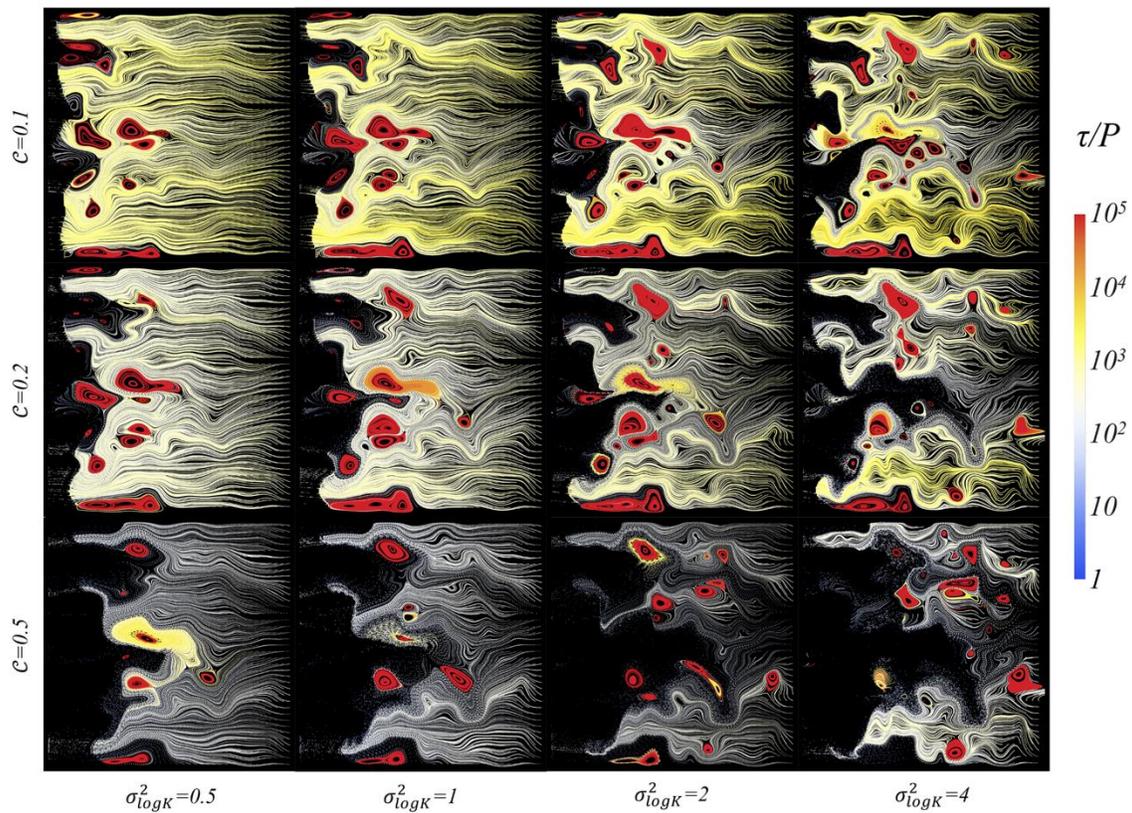

**Figure S3**. Poincaré sections (distributed injection) of flow in a single log-$K$ realization for $\mathcal{T} = 10$ and $\mathcal{G} = 100$ with varying $\mathcal{C}$ and $\sigma_{\log K}^2$. The characteristics of the heterogeneous field: $K_{\text{eff}} = 2 \times 10^{-4}$, $\lambda = 0.049$.

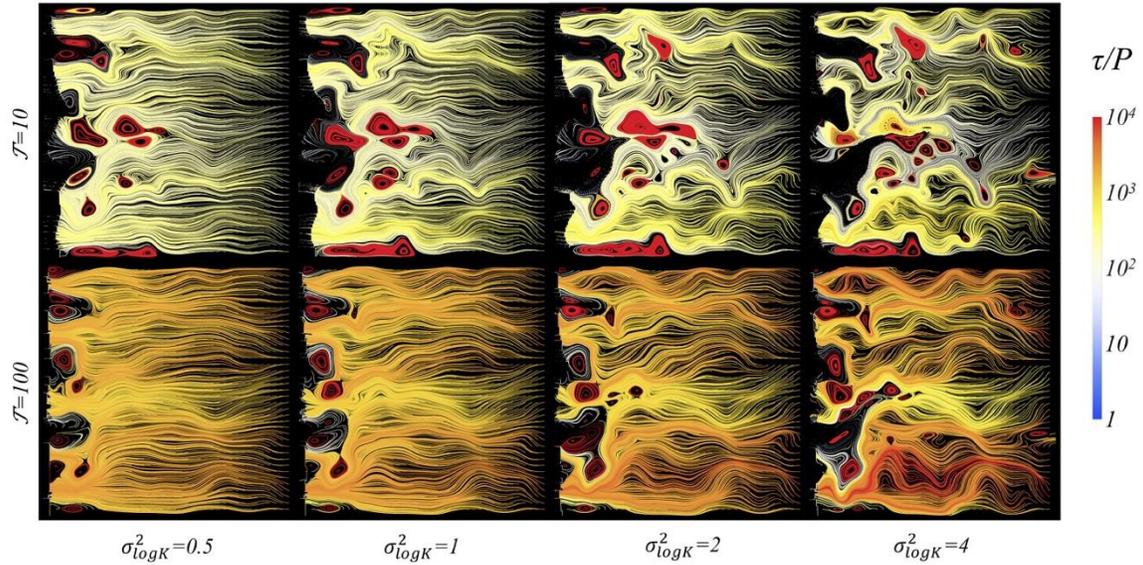

**Figure S4.** Poincaré sections (distributed injection) of flow in a single log-$K$ realization for $\mathcal{C} = 0.1$ and $\mathcal{G}$ = 100 with varying $\mathcal{T}$ and $\sigma^2_{\log K}$. The characteristics of the heterogeneous field: $K_{\mathrm{eff}} = 2 \times 10^{-4}$, $\lambda = 0.049$.

**Text S4. Impact of correlation length**

Figure S5 illustrates the effect of different correlation lengths ($\lambda$ = 0.025, 0.049, 0.074, 0.098) upon the aquifer transport structures for the fixed dynamical parameters $\mathcal{T}$ = 10 and 100, $\mathcal{G}$ = 100, $\mathcal{C}$ = 0.1 and log-conductivity variance $\sigma^2_{\log K}$ = 2. As expected, the size and extent of the resulting closed flow regions are controlled by the correlation length in that as $\lambda$ increases, fewer and larger closed flow regions occur in the domain. These results indicate that the correlation length is a key aquifer property that controls the size of closed regions within the flow, and in Section 5.2 of the manuscript, we show that this size is important in the transport of dispersive solutes.

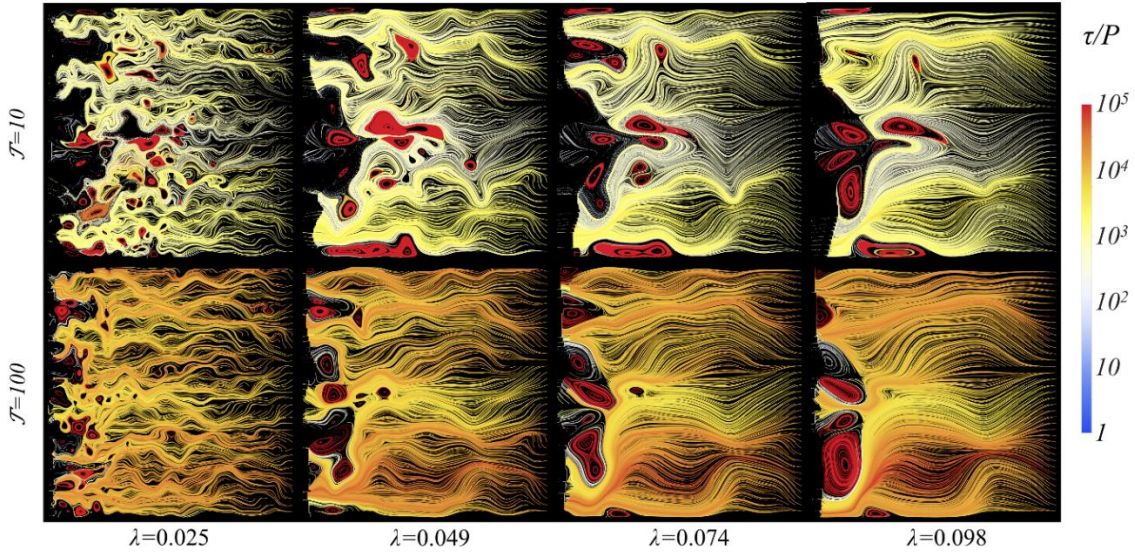

**Figure S5**. Poincaré sections (distributed injection) for $\mathcal{G} = 100$ and $\mathcal{C} = 0.1$ with varying $\mathcal{T}$ and $\lambda$. The characteristics of the heterogeneous field: $K_{\text{eff}} = 2 \times 10^{-4}$, $\sigma^2_{\log \kappa} = 2$.

Figure S6 shows the average radius of trapped regions for two cases where $\mathcal{G} = 100$, $\mathcal{C} = 0.1$ and $\mathcal{T}$ varies between 10 and 100 against a range of correlation lengths ($\lambda = 0.025, 0.049, 0.074, 0.098$). For both cases the mean radius increases linearly with correlation length $\lambda$ and this rate of increase also decreases with Townley number $\mathcal{T}$. This is expected as decreasing $\mathcal{T}$ corresponds to increasing width of the tidal region and a larger number of trapped regions in the aquifer domain.

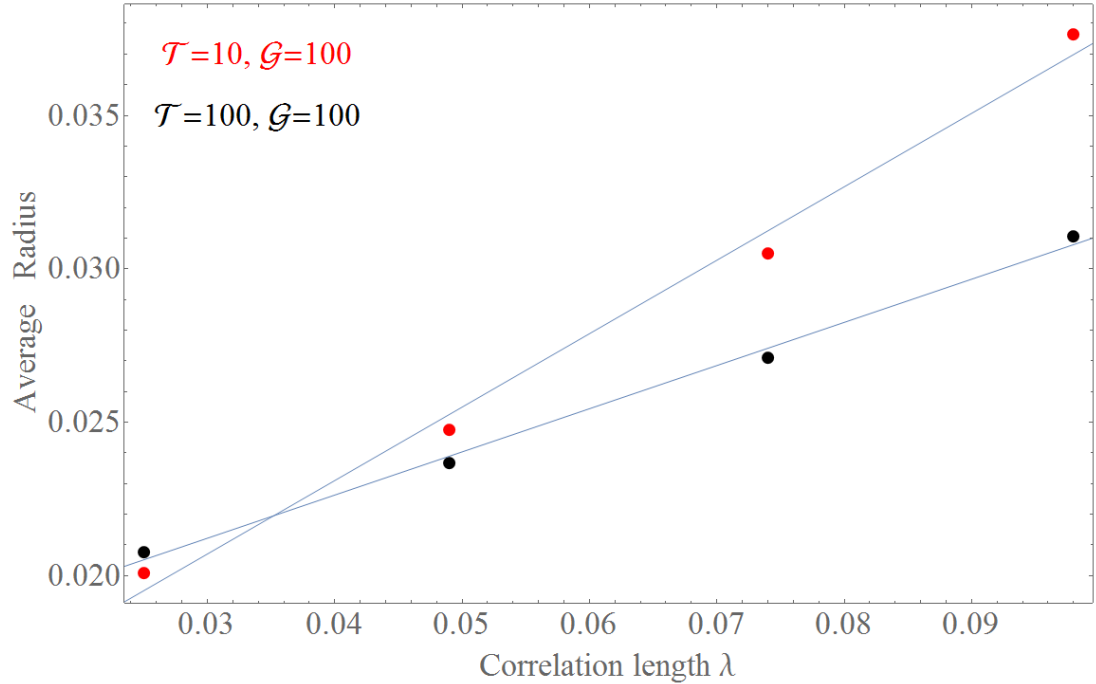

**Figure S6.** Average radius of trapped regions in the aquifer for different values of the correlation length $\lambda$ for two cases ($\mathcal{T} = 10$ and $100$) with fixed tidal strength and compressibility ($\mathcal{G} = 100$, $\mathcal{C} = 0.1$).

In TLMW, we constructed numerical methods to generate self-consistent numerical solutions to physically constrained flows and accurate advection methods to construct Poincaré sections. It was shown that the accuracy of the method for generating physically constrained flows declines as log-conductivity variance $\sigma_{\log K}^2$ increases and as the spatial resolution $\lambda/\Delta$ is reduced (where $\Delta$ is the spatial resolution of the finite difference grid). Thus, while the sections in Figure S2 show a trend of increasing structure and complexity as $\lambda/\Delta$ *decreases*, the accuracy of the underlying flow fields increases as $\lambda/\Delta$ *increases*. Even so, it is clear from the figure that the calculated time-averaged particle trajectories are smoothly varying for increasing $\lambda/\Delta$.

**Text S5. Distributions of $R_0^2$ and $R_0^2/v$**

In Section 5.2 of the manuscript, we showed that the residence time of diffusive and dispersive solutes within trapped regions of the flow respectively scales roughly as the squared relative radius ($R_0^2$) of the trapped region, and this area normalized by the local fluid velocity ($R_0^2/v$). Figure S7 (a) and (b) shows the distribution of $R_0^2$ and $R_0^2/v$ for various values of the Townley number and tidal strength parameters, indicating that larger closed regions and hence slower transport of diffusive solutes occurs for large values of $\mathcal{G}$, whereas smaller values of results in smaller closed regions and slower transport.

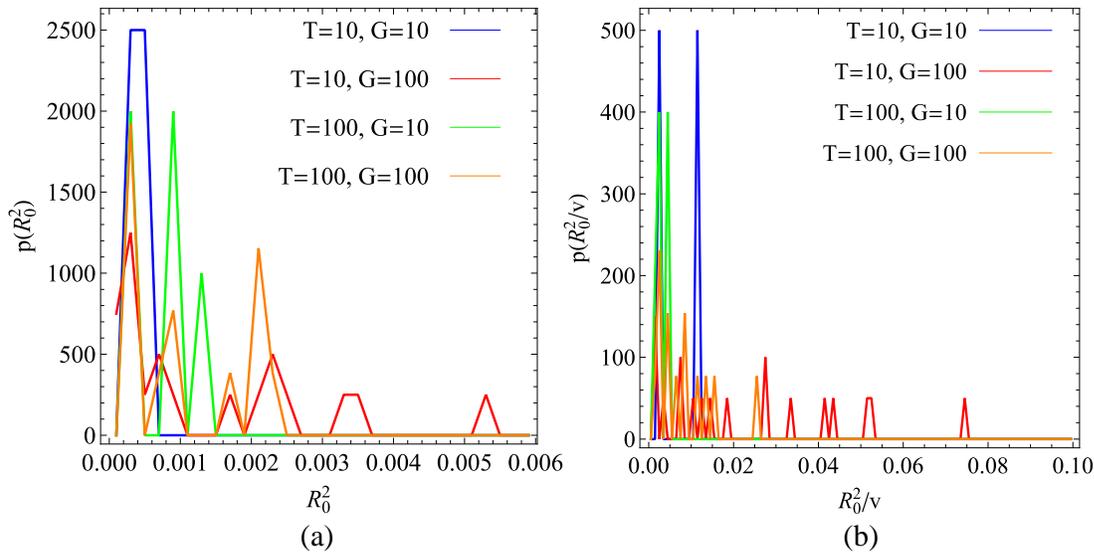

**Figure S7.** Histograms (probability density function) of (a) $R_0^2$ and (b) $R_0^2/v$ which govern solute transport across trapped regions respectively via molecular diffusion and hydrodynamic dispersion for Townley numbers and tidal strengths $\mathcal{T}$ =10, $\mathcal{G}$ =10 (blue), $\mathcal{T}$ =10, $\mathcal{G}$ =100 (red) , $\mathcal{T}$ =100, $\mathcal{G}$ =10 (green), $\mathcal{T}$ =100, $\mathcal{G}$ =100 (orange).

**Text S6. Mean velocity variation**

In Section 5.2 we relate the residence time of dispersive solutes to the size of trapped regions and the local fluid velocity. To provide an estimate of this residence time, we need to relate the mean regional flow velocity to the tidally influenced velocity near the tidal boundary. The mean regional groundwater velocity is

$$\bar{v}_r = -J \frac{K_{\text{eff}}}{\varphi_{\text{eff}}}, \tag{S2}$$

and the maximum tidal groundwater velocity $\tilde{v}_{max}$ can be shown to be

$$\tilde{v}_{max} = \bar{v}_r \left(1 + \frac{\mathcal{G}}{(1 - \mathcal{C}) x_{\text{taz}}}\right). \tag{S3}$$

We consider some field-scale parameters pertinent to a Safety Bay Sand/Tamala Limestone coastal aquifer near Fremantle (Smith et al., 2012). Representative parameter values are: aquifer width $L \approx 2000$ m, dimensionless head gradient $J \approx 0.005$, hydraulic conductivity $K_{\text{eff}} \approx 10$ m/d, dimensionless porosity $\varphi_{\text{eff}} \approx 0.3$, dimensionless tidal strength in Fremantle $\mathcal{G} \approx 0.5 / (JL)$, tidally affected zone (dimensionless fraction of $L$) $x_{\text{taz}} \approx 0.2$, dimensionless tidal compressibility $\mathcal{C} \approx 0.03$. Numerical evaluations show that the regional flow velocity $\bar{v}_r$ is well within an order of magnitude estimator of the maximum tidally induced velocity $\tilde{v}_{max}$ for practical purposes. But for extreme values of $\mathcal{G}$ and $\mathcal{C}$, the tidally induced velocity becomes significantly greater than the regional velocity.

### Text S7. Relative penetration distance of saline wedges

In coastal aquifers, the presence of a saltwater/freshwater interface (which often manifests as a saline wedge) can generate significant density-driven flows that are not accounted for in our current model. However, these flows are often confined to a small zone close to the coastal boundary, whereas the coastal aquifer may extend much further inland. To estimate the magnitude of the impact of these density currents upon our model predictions, below we employ a sharp interface model (Strack, 1976) to calculate the penetration distance of the saline wedge toe ($x_{\text{wt}}$) relative to that of the tidal signal ($x_{\text{taz}}$) in coastal aquifers. Due to neglect of dispersion the sharp interface model will overestimate wedge penetration and thus will provide a conservative measure of the spatial extent of density currents near the discharge boundary. A non-dimensional expression for $x_{\text{taz}}$ can be found in TLMW. For a homogeneous aquifer the penetration distance of the saline wedge toe in the presence of vertical recharge is given by (Somaratne & Ashman, 2018) as

$$x_{\text{wt}} = \frac{q}{R} + L - \sqrt{\left(\frac{q}{R} + L\right)^2 - \frac{K}{R}\delta(1 + \delta)d^2},\qquad\text{(S4)}$$

where $q \equiv -K\,J$ is the regional flux (with conductivity $K$ and regional gradient $J$), $R$ the vertical recharge rate, $d$ the aquifer thickness, and $\delta \equiv (\rho_s - \rho_f)/\rho_f \approx 0.025$ is the salinity ratio (where $\rho_s, \rho_f$ are the saline and freshwater densities, respectively).As the wedge toe is located where the saline wedge reaches zero thickness, we define the ratio of the half-wedge position $x_{\text{wt}}/2$ to $x_{\text{taz}}$ as the *wedge ratio* W. In this way W measures the position where the saline interval of the water column equals the freshwater interval. Figure S8 plots W as a function of regional flux for a range of different values of aquifer thickness, storativity and conductivity that are representative of coastal aquifers. Somewhat counter-intuitively, W increases with aquifer conductivity $K$ as the associated increase in wedge toe position $x_{\text{wt}}$ dominates over the associated increase in $x_{\text{taz}}$. Similarly, an increase in the regional gradient $J$ counter-intuitively increases W as the increase in $J$ decreases the tidally active zone (and hence $x_{\text{taz}}$) by a greater extent than it does the wedge toe position $x_{\text{wt}}$. As expected, increasing aquifer compressibility $S$ acts to decrease penetration of tidal signals and thus increase W.

For common regional gradients ($J < 0.01$), storativities ($S < 0.01$) and conductivities ($K < 50$ m/d), W is significantly smaller than unity, hence there exist significant non-saline intervals in the saturated thickness of the aquifer through which tidal signals can propagate according to standard freshwater Darcian dynamics, Reducing $K$ and/or increasing $g_p$ lead to further reductions in W. Thus, for a broad range of physical aquifer parameters there remains a significant portion of the coastal aquifer systems that is not impacted by density currents but is still subject to tidal forcing.

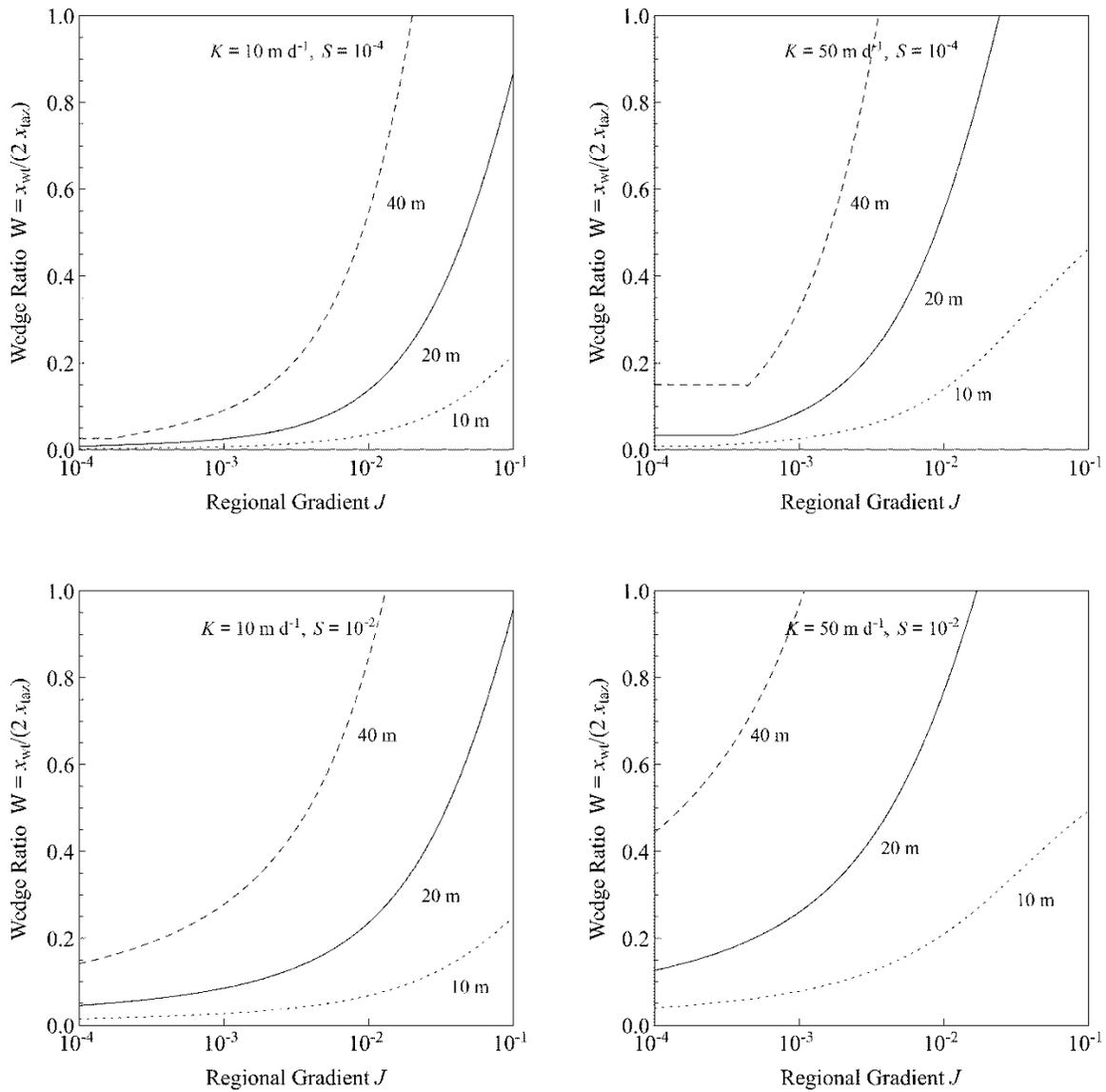

**Figure S8.** Estimates of the wedge ratio W as a function of regional gradient *J* in moderate conductivity (*K* = 10 m/d, left column) and high conductivity (*K* = 50 m/d, right column) coastal aquifers for various values of aquifer thickness *d* and storativity *S*. Recharge *R* and tidal amplitude $g_p$ are held fixed at 0.001 m d$^{-1}$ and 1 m, respectively.

**Text S8. Transition of aquifer transport from open to closed transport structures**

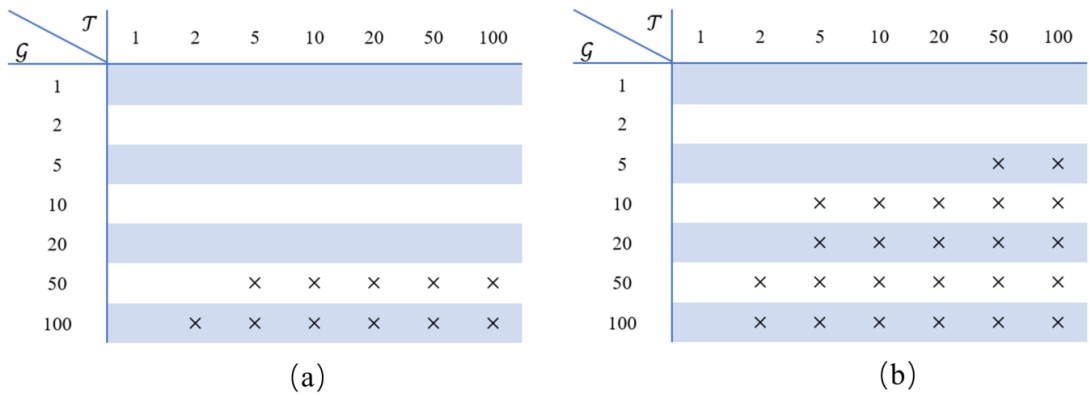

**Figure S9.** Summary of transport simulations that exhibit closed transport structures (as indicated by crosses) for compressibility ratio (a) $\mathcal{C} = 0.01$ and (b) $\mathcal{C} = 0.1$.

Figure S9 above summarizes the transport simulations that exhibit closed transport structures, indicating that the propensity for complex transport increases with compressibility ratio $\mathcal{C}$ and Townley number $\mathcal{T}$.

**Text S9. Tracer residence times and size of trapped regions**

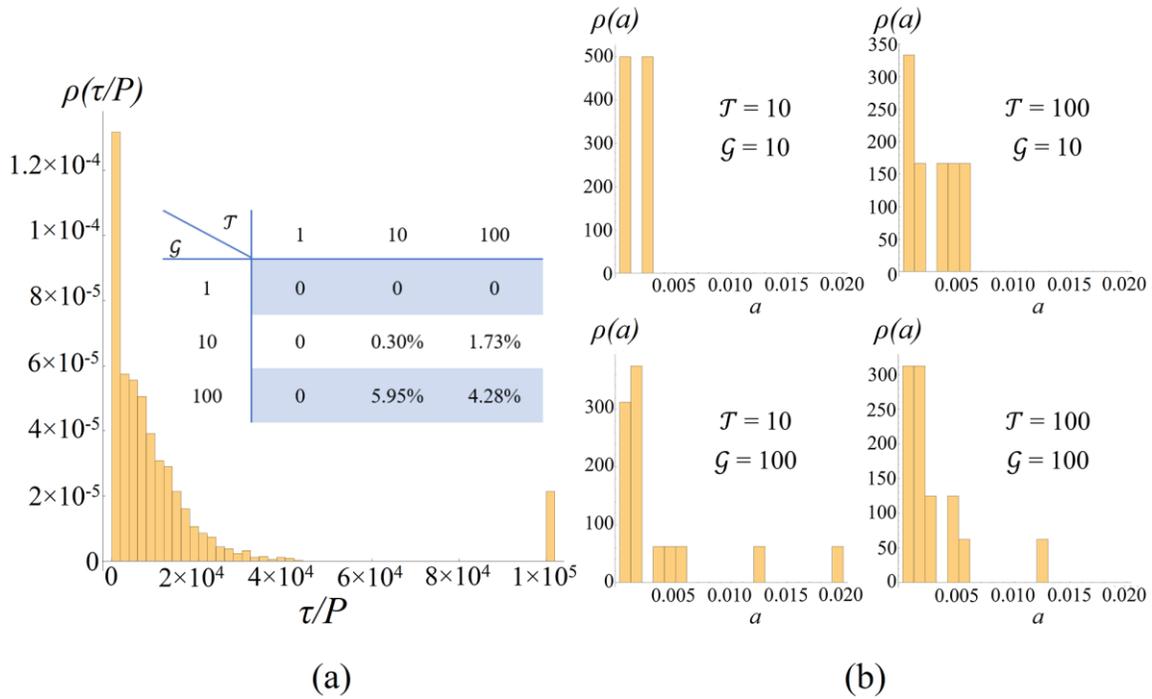

(a)                                   (b)

**Figure S10.** (a) Residence time distribution of fluid particles for the distributed injection protocol for $\mathcal{T} = 100$, $\mathcal{G} = 100$ at $\mathcal{C} = 0.1$. Inset: total of the area (*a*) of trapped regions (closed orbits) as a proportion of aquifer area for various values of $\mathcal{T}$ and $\mathcal{G}$ at $\mathcal{C} = 0.1$. (b) Histograms (probability density function) of the relative area (*a*) of individual trapped regions in the aquifer for various values of $\mathcal{T}$ and $\mathcal{G}$ at $\mathcal{C} = 0.1$.

Figure S10a indicates the residence time distribution of tracer particles in the aquifer model for $\mathcal{T} = 100$, $\mathcal{G} = 100$ at $\mathcal{C} = 0.1$. This plot shows that many particles exit within a few thousand flow periods, followed by a long tail characteristic of transport in heterogeneous media. In addition, a significant portion of the particles remain trapped in closed regions, as indicated by the tracer particles with residence times equivalent to that of the simulation time of $10^5$ forcing periods. The inset of Figure S10a indicates the relative total area (a) of trapped regions in the aquifer increases with both $\mathcal{T}$ and $\mathcal{G}$ at $\mathcal{C} = 0.1$, and this trend has been observed to persist for other values of $\mathcal{C}$. Figure 10b summarises the distribution of the relative size of individual trapped regions for various values of $\mathcal{T}$ and $\mathcal{G}$ at $\mathcal{C} = 0.1$, indicating the size of individual trapped regions tends to increase with $\mathcal{T}$ and $\mathcal{G}$.

**References**

Smith, A., Massuel, S., Pollock, D., & Dillon, P. (2012). Geohydrology of the Tamala Limestone Formation in the Perth Region: Origin and role of secondary porosity. https://doi.org/10.4225/08/599dd0fd98a44

Somaratne, N., & Ashman, G. (2018). Analysis of Saline Intrusion into a coastal Aquifer: A Case History of Legacy Issues and Challenges to water security. *Environ Nat Resour Res, 8*(2), 16-32. doi:https://doi.org/10.5539/enrr.v8n2p16

Strack, O. D. L. (1976). A single-potential solution for regional interface problems in coastal aquifers. *Water Resources Research, 12*(6), 1165-1174. doi:https://doi.org/10.1029/WR012i006p01165